\documentclass[sn-mathphys]{sn-jnl}

\usepackage{pifont}
\usepackage{colortbl}

\jyear{2021}%

\theoremstyle{thmstyleone}%
\theoremstyle{thmstyletwo}%

\theoremstyle{thmstylethree}%

\newcommand{\fref}[1]{Figure~\ref{#1}}
\newcommand{\tref}[1]{Table~\ref{#1}}

\newcommand{\aref}[1]{Appendix~\ref{#1}}

\begin{document}

\title[Homeostatic regulation of renewing tissue cell populations via crowding control]{Homeostatic regulation of renewing tissue cell populations via crowding control: stability, robustness and quasi-dedifferentiation}


\author[1,2,3]{\fnm{Cristina} \sur{Parigini}}\email{cristina.parigini@auckland.ac.nz}

\author*[1,2]{\fnm{Philip} \sur{Greulich}}\email{p.s.greulich@soton.ac.uk}

\affil*[1]{\orgdiv{School of Mathematical Sciences}, \orgname{University of Southampton}, \city{Southampton}, \country{United Kingdom}}

\affil[2]{\orgdiv{Institute for Life Sciences}, \orgname{University of Southampton}, \city{Southampton}, \country{United Kingdom}}

\affil[3]{\orgdiv{Te P\=unaha \=Atea -  Space Institute}, \orgname{University of Auckland}, \city{Auckland}, \country{New Zealand}}



\abstract{To maintain renewing epithelial tissues in a healthy, homeostatic state, (stem) cell divisions and differentiation need to be tightly regulated. Mechanisms of homeostatic control often rely on crowding control: cells are able to sense the cell density in their environment (via various molecular and mechanosensing pathways) and respond by adjusting division, differentiation, and cell state transitions appropriately. Here we determine, via a mathematically rigorous framework, which general conditions for the crowding feedback regulation (i) must be minimally met, and (ii) are sufficient, to allow the maintenance of homeostasis in renewing tissues. We show that those conditions naturally allow for a degree of robustness toward disruption of regulation. Furthermore, intrinsic to this feedback regulation is that stem cell identity is established collectively by the cell population, not by individual cells, which implies the possibility of `quasi-dedifferentiation', in which cells committed to differentiation may reacquire stem cell properties upon depletion of the stem cell pool. These findings can guide future experimental campaigns to identify specific crowding feedback mechanisms.
}




\maketitle

\section{Introduction}\label{sec:intro}

Many adult tissues are {\em renewing}, that is, terminally differentiated cells are steadily removed and replaced by new cells produced by the division of cycling cells (stem cells and progenitor cells), which then differentiate. In order to maintain those tissues in a healthy, homeostatic state, (stem) cell divisions and differentiation must be tightly balanced. Adult stem cells are the key players in maintaining and renewing such tissues due to their ability to produce cells through cell division and differentiation persistently \citep{NIHStemCell2016}. However, the underlying cell-intrinsic and extrinsic factors that regulate a homeostatic state are complex and not always well understood.
 
 Several experimental studies have identified mechanisms and pathways that regulate homeostasis. For example, cell crowding can trigger delamination and thus loss of cells in Drosophila back  \citep{Marinari2012Live-cellOvercrowding}, and differentiation in cultured human colon, various zebrafish epiderimises, and canine kidney cells \citep{Eisenhoffer2012CrowdingEpithelia, Eisenhoffer2013BringingNumbers}. On the other hand, cell crowding can affect cell proliferation: overcrowding can inhibit proliferation \citep{Puliafito2012}, whereas a reduction in the cell density, obtained, for example, by stretching a tissue \citep{Gudipaty2017} causes an increase in proliferative activity (both shown in cultured canine kidney cells). Although the mechanisms to mediate this regulation are not always clear, experimental studies on mechanosensing showed that cell overcrowding reduces cell motility and consequently produces a compression on cells that inhibits cell proliferation \citep{Puliafito2012,Shraiman2005MechanicalGrowth}. Another mechanism utilising crowding feedback is the competition for limited growth signalling factors \citep{Kitadate2019CompetitionNiche}. More specifically, in the mouse germ line, cells in the niche respond to a growth factor (FGF5) that promotes proliferation over differentiation, which they deplete upon being exposed to it. Therefore, the more cells are in the niche, the less FGF5 is available per cell, and the less proliferation (or more differentiation) occurs. 
 
 Despite differing in the involved molecular pathways and many other details, all these regulatory mechanisms are, in essence, sensing the cell density in their environment and responding by adjusting their propensities to divide, differentiate, die, or emigrate from the tissue. This class of mechanisms, for which cell fate propensities depend on the cell density, can be classified as \emph{crowding feedback} regulation: the cell density determines the cells' proliferation and differentiation, which affects their population dynamics and thus the cell density. However, the crowding response to changes in cell density cannot be arbitrary in order to maintain homeostasis. It must provide a (negative) \emph{feedback}, in the sense that cells sense the cell density and adjust proliferation, differentiation, and cell loss, such that the cell density is decreased if it is too high and increased if it is too low. For simple tissues consisting of a single cell type with a unique cell state, it is relatively straightforward to give the conditions for crowding feedback to maintain homeostasis successfully. In this case, when the cell division rate decreases with cell density and differentiation and or death rate increase with cell density, a homeostatic state is maintained. However, such conclusions are not as simple to make when a tissue consists of a complex lineage hierarchy and a multitude of underlying cellular states. In the latter, more realistic case, conditions for successful homeostatic regulation -- in which case we speak of \emph{crowding control} -- may take more complex forms. 
 
Previous studies based on mathematical modelling have shed some light on quantitative mechanisms for homeostatic control \citep{Johnston2007,Sun2012StochasticZheng, Bocharov2011FeedbackNumber,Greulich2016,Greulich2021}. In particular, in \citep{Greulich2021}, a mathematical assessment of crowding feedback modelling shows that a (dynamic) homeostatic state exists under reasonable biological conditions. Nevertheless, the case of dynamic homeostasis considered there may not necessarily be a steady state but could also exhibit oscillations in cell numbers (as does realistically happen in the uterus during the menstrual cycle). While the criterion presented in \citep{Greulich2021} provides a valid sufficient condition for dynamic homeostasis, it relies on a rather abstract mathematical quantity -- the dominant eigenvalue of the dynamical matrix -- that is difficult, if not impossible, to measure in reality.
 
Here, we wish to generalise previous findings and seek to identify general conditions for successful homeostatic control if propensities for cell division, differentiation, and loss are responsive to variations in cell density. More precisely, we derive conditions that must be minimally fulfilled (necessary conditions) and conditions which are sufficient, to ensure that homeostasis prevails. To identify and formulate those conditions, we note that homeostasis is a property of the tissue cell population dynamics, which can be mathematically expressed as a dynamical system. Even if a numerically exact formulation of the dynamics may not be possible, one can formulate generic yet mathematically rigorous conditions by referring to the criteria for the existence of stable steady states in the cell population dynamics of renewing tissues. We will derive those conditions by mathematical, analytical means, augmented by a numerical analysis testing the limits of those conditions.
 
We will also show that homeostatic control by crowding feedback possesses inherent robustness to failures and perturbations of the regulatory pathways, which may occur through external influences (e.g. wide-spread biochemical factors) and genetic mutations. Finally, we will assess the response of cells when the pool of stem cells is depleted. Crucially, we find that inherent to crowding feedback control is that formerly committed progenitor cells reacquire self-renewal capacity without substantial changes in their internal states. Dedifferentiation has been widely reported under conditions of tissue regeneration \citep{DonatiWatt2017,Jopling2011} or when stem cells are depleted \citep{Tata2013DedifferentiationVivo,Tetteh2015,Tetteh2016,Murata2020}, which is usually thought to involve a substantial reprogramming of the cell-intrinsic states towards a stem cell type. On the other hand, our analysis suggests the possibility of ``quasi''-dedifferentiation, the reversion from a committed cell to a stem cell by a mere quantitative adjustment of the pacing of proliferation and differentiation, without a substantial qualitative change in its expression profiles. 

\section{Modelling of tissue cell dynamics under crowding feedback}\label{sec:modelling}

We seek to assess the conditions for homeostasis in renewing tissue cell populations, that is, either a steady state of the tissue cell population (strict homeostasis) or long-term, bounded oscillations or fluctuations (dynamic homeostasis), which represent well-defined constraints on the dynamics of the tissue cell population. To this end, we will here derive a formal, mathematical representation of the tissue cell dynamics under crowding feedback. 

The cell population is fully defined by (i) the number of cells, (ii) the internal (biochemical and mechanical) states of each cell, and (iii) the spatial position of cells. We assume that a cell's behaviour can depend on the cell density and the states of cells in its close cellular environment. As we examine a situation close to a homeostatic state, we assume that the cell density is homogeneous over the range of interaction between cells, which expands over a volume $V$. Hence, the cell density $\rho$ is proportional to the average number of cells, $\bar n$, in that volume, $\rho = \frac{\bar n}{V}$. Similarly, we define the number of cells in internal state $i$ as $n_i$, and the cell density of cells in internal state $i$ as $\rho_i = \frac{\bar n_i}{V}$, where $\bar n_i$ is the expected value of $n_i$. As we consider only the crowding feedback response of cells, which only accounts for the cell densities $\rho_i$ but not the explicit position of cells, the spatial configuration (iii) is not relevant to our considerations. Thus, the configuration of the cell population and its time evolution is entirely determined by the average number of cells in each state $i$, as a function of time $t$, $\bar n_i(t)$. The configuration of cell numbers $n_i$ can change only through three processes: (1) cell division, whereby it must be distinguished between the cell state of daughter cells, (2) the transition from one cell state to another, (3) loss of a cell, through cell death or emigration out of the tissue. Following the lines of Refs. \citep{Greulich2021,Parigini2020UniversalityStrategies} and denoting as $X_{i,j,k}$ a cell in internal states $i,j,k$, respectively, we can formalise these events as: 
\begin{align}
\mbox{ cell division: } X_i &\xrightarrow{\lambda_i r_i^{jk}} X_j + X_k 
\label{mdl:GenericProcessDivision} \\
\mbox{ cell state transition: } X_i &\xrightarrow{\omega_{ij}} X_j  \label{mdl:GenericProcessTransition} \\
\mbox{ cell loss: } X_i &\xrightarrow{\gamma_i}\emptyset \,\,\ , \label{mdl:GenericProcessDeath}
\end{align}
where the symbols above the arrows denote the dynamical rates of the transitions, i.e. the average frequency at which such events occur. In particular, $\gamma_i$ is the rate at which a cell in state $i$ is lost, $\omega_{ij}$ the rate at which a cell changes its state from $i$ to $j$ and $\lambda_i r_i^{jk}$ denotes the rate at which a cell $i$ divides to produce two daughter cells, one in state $j$ and one in state $k$ ($i=j,j=k,k=i$ are possible). For later convenience, we distinguish here the overall rate of cell division in state $i$, $\lambda_i$ and the probability $r_i^{jk}$ that such a division produces daughter cells in states $j$ and $k$. 

Since we consider a situation where cells can respond to the cell densities $\rho_i$ via crowding feedback, all the rates and probabilities ($\lambda_i,\gamma_i,\omega_{ij},r_i^{jk}$) may depend on the cell densities of either state $j$, $\rho_j$. For convenience, we discretise the number of states in case the state space is a continuum and only distinguish states which have substantially different propensities ($\lambda_i,\gamma_i,\omega_{ij},r_i^{jk})$. Without loss of generality, we assume that there are $m$ states, that is, $i,j,k=1,...,m$ (for a rigorous argument for the discretisation of the state space, see \citep{Greulich2021}).

The rates given above denote the \emph{average} number of events happening per time unit. Thus, we can express the total rate of change of the \emph{average} (i.e. expected) number of cells $\bar n_i(t)$, that is, the derivative $\dot{\bar n}_i = \frac{d\bar n_i}{dt}$, in terms of the rates of those events. This defines a set of ordinary differential equations.
Following the lines of Refs. \citep{Greulich2021,Parigini2020UniversalityStrategies}, we can write $\dot n_i$ as,
\begin{align}
\label{dyn-eq1}
    \dot{\bar n}_i = \left[\sum_j \omega_{ji} \bar n_j +  \lambda_j \left(\sum_k r_j^{ik} + r_j^{ki}\right) \bar n_j\right] - \bar n_i \left(\lambda_i + \gamma_i + \sum_j \omega_{ij}\right) \,\,\ ,
\end{align}
where for convenience, we did not write the time dependence explicitly, i.e. $n_i = n_i(t)$, and all parameters may depend on the cell densities $\rho_j$. Since $V$ is constant, we can divide by $V$ to equivalently express this in terms of the cell state densities, $\rho_i = \frac{\bar n_i}{V}$, and then write Eq. \eqref{dyn-eq1} compactly as,
\begin{align}
\label{dyn_sys_feedback_eq}
\frac{d}{dt}\bm \rho(t) = A(\bm \rho(t)) \, \bm \rho(t)
\end{align}
where $\bm \rho = (\rho_1,\rho_2,...)$ is the vector of cell state densities and $A(\bm \rho)$ is the matrix,
\begin{align} \label{eq:AdjacencyMatrixDef}
    A = \begin{pmatrix} 
    \lambda_1 - \sum_{j \neq 1} \kappa_{1j} - \gamma_1 & \kappa_{21} & \kappa_{31} & \cdots \\
    \kappa_{12} & \lambda_2  - \sum_{j \neq 2} \kappa_{2j} - \gamma_2 & \kappa_{32} & \cdots \\
    \kappa_{1m} & \kappa_{2m} & \cdots & \lambda_m - \sum_{j \neq m} \kappa_{mj} - \gamma_m
    \end{pmatrix} \, ,
\end{align}
in which $\kappa_{ij} = \lambda_i 2 r_i^j + \omega_{ij}$, with $r_i^j = \sum_k (r_i^{jk}+r_i^{kj})/2$, is the total transition rate, that combines all transitions from $X_i$ to $X_j$ by cell divisions and direct state transitions (again, all parameters may depend on $\bm \rho$, as therefore also does $A$). We can thus generally write the elements of the matrix $A$, $a_{ij}$ with $i,j=1,...,m$ as
\begin{align}
\label{eq:MatrixElements}
a_{ij} = \left\lbrace \begin{array}{cc}
  \lambda_i - \gamma_i - \sum_{k \neq i} \kappa_{ik}   & \mbox{ for } i = j  \\
   \kappa_{ji}  & \mbox{ for } i \neq j
\end{array}\right.
\end{align}

We now make the mild assumption that divisions of the form $X_i \to X_j + X_k$ are effectively three events, namely, cell duplication, $X_i \to X_i + X_i$ coupled to cell state changes, $X_i \to X_j$ and $X_i \to X_k$, if $j \neq i$ or $k \neq i$. In this view, the parameters relevant for crowding feedback are the total cell state transition propensities $\kappa_{ij}$ and the cell division rate $\lambda_i$, as in \eqref{eq:AdjacencyMatrixDef}, instead of $\omega_{ij}$ and $r_i^{jk}$.

These equations describe a dynamical system which, for given initial conditions, determines the time evolution of the cell densities, $\rho_i(t)$. Crucially, this description allows for a rigorous mathematical definition of what a homeostatic state is, and to apply tools of dynamical systems analysis to determine the circumstances under which a homeostatic state prevails. In particular, we define a \emph{strict} homeostatic state as a steady state of the system, \eqref{dyn_sys_feedback_eq}, when the cell numbers -- and thus cell densities, given that $V$ is fixed -- in each state do not change, mathematically expressed as $\frac{d{\bm \rho}}{dt} = 0$ (a fixed point of the system). A dynamic homeostatic state is when cell densities may also oscillate or fluctuate but remain bounded and thus possess a finite long-term average cell population (in which case the system either approaches a steady state or limit cycles -- that is, oscillations -- or chaotic but bounded behaviour). Based on these definitions, we can now analyse under which circumstances crowding feedback can maintain those states, which in the case of strict homeostasis requires, in addition, that the corresponding steady state is stable.

\subsection{Cell types and lineage hierarchies}
\label{sec:CellTypes}
According to \citep{Greulich2021}, cell population dynamics of the type \eqref{dyn_sys_feedback_eq} can be associated with a cell state network, in which each state is a node, and the nodes are connected through cell state transition (direct transitions and cell divisions). Furthermore, by decomposing this network in \emph{strongly connected components (SCCs)}, the cell fate model can be viewed as a directed acyclic network \citep{Bollobas1998Modern}, generally called the \emph{condensed network}. Here, we follow the definitions of \citep{Greulich2021} and define a \emph{cell type} as an SCC of the cell state network, so that any cell states connected via cyclic cell state trajectories (sequences of cell state transitions) are of the same type, and the condensed network of cell types represents the \emph{cell lineage hierarchy}. This definition ensures that cells of the same type have the same lineage potential (outgoing cell state trajectories) and that the stages of the cell cycle are associated with the same cell type. In this context, we will in the following also speak of \emph{differentiation} when a cell state transition between different cell types occurs. 

Each cell type can be classified as self-renewing, declining or hyper-proliferating, depending on the dominant eigenvalue $\mu$ (called \emph{growth parameter}) of the dynamical matrix $A$ (from Eq. \eqref{dyn_sys_feedback_eq} ff.)  reduced to that SCC. This is $\mu = 0$ for self-renewing cell types, when cell numbers of that type remain constant over time, $\mu < 0$ ($\mu > 0$) for the declining (hyperproliferating) types when cell numbers decline (increase) in the long term \citep{Greulich2021}. Importantly, for the population dynamics to be strictly homeostatic, which means that a steady state of model \eqref{dyn_sys_feedback_eq} exists, the cell type network must fulfil strict rules. These are: (i) at least one self-renewing cell type (with $\mu=0$) must exist; (ii) self-renewing cell types must stay at an apex of the condensed network; (iii) all the other cells must be of declining types. This means that the critical task of homeostatic control is to ensure that the kinetic parameters of the cell type at the apex of the cell lineage hierarchy are fine-tuned to maintain exactly $\mu=0$. 

Therefore, we can restrict our analysis to find conditions for the cell type at the lineage hierarchy's apex to be self-renewing, which we will do in the following. Other cell types simply need that differentiation (transition towards another cell type) or loss is faster than proliferation, so that they become declining cell types, $\mu < 0$, but those rates do not require fine-tuning and thus trivially regulated. We note that when we consider only cell states of the type at the apex of the cell lineage hierarchy, any differentiation event is -- according to this restricted model -- a cell loss event and included as event occurring with rates $\gamma_i$. Given that cell loss from a cell type at the lineage apex is rare, we will therefore in the following also denote the rates $\gamma_i$ simply as \emph{differentiation rates}.

\section{Results}\label{sec:Results}

We will now determine necessary and sufficient conditions for the establishment of strict and dynamical homeostasis when subject to crowding feedback, which we here define through the derivatives of the dynamical parameters $\lambda_i,r_i^{jk},\omega_{ij}, \gamma_i$ as a function of the cell densities. As argued before, we only need to consider cell types at an apex of the cell type network, which, for homeostasis to prevail, must have a growth parameter (i.e. dominant eigenvalue of matrix $A$ in Eq. \eqref{eq:AdjacencyMatrixDef}) $\mu=0$. Furthermore, we assume that the apex cell type resides in a separate stem cell niche. Therefore, the parameters only depend on cell densities $\rho_i$ of states associated with that cell type, i.e. we can write $A = A(\rho)$, where $\rho = \sum_{i\in S} \rho_i$ comprises only cell states of the apex cell type $S$. Provided that, the matrix elements are functions of $\rho$, and therefore also $\mu$ is a function of $\rho$. Thus, self-renewal corresponds to a non-trivial fixed point, $\bm \rho^*$, of Eq. \eqref{dyn_sys_feedback_eq}, restricted to cell type $S$, for which the dominant eigenvalue of $A$ is zero, that is $\mu(\rho^*) = 0$ ($\rho^* = \sum_{i\in S} \rho^*_i$).

For convenience, we will often generally refer to parameters as $\alpha_i$, $i=1,...,2m+m^2$, where $\alpha_i$ stands for any of the parameters, $\lbrace \lambda_i,\gamma_i,\kappa_{ij}\vert i,j=1,...,m \rbrace$, respectively\footnote{More precisely, $\alpha_i\vert_{i=1,..,m} := \lambda_i$, $\alpha_i\vert_{i=m+1,..,2m} := \gamma_{i-m}$, $\alpha_i\vert_{i=2m+1,..,2m+m^2} := \kappa_{\lfloor (i-2m)/m \rfloor,i-\lfloor (i-2m)/m \rfloor m}$}. Hence, we study which conditions the functions $\alpha_i(\rho)$ must meet to maintain homeostasis. In particular, we study how those parameters qualitatively change with the cell density -- increase or decrease -- that is, we study how the sign and magnitude of derivatives $\alpha_i' :=\frac{d\alpha_i}{d\rho}$ affects homeostasis.

A crucial property of the matrix $A(\rho)$ is that it is always a \emph{Metzler matrix}, since all its off-diagonal elements, $\kappa_{ij} \geq 0$. Since the cell state network of a cell type is strongly connected, we can further state that $A(\rho)$ is irreducible. Notably, for irreducible Metzler matrices holds the Perron-Frobenius theorem \citep{MacCluer2000TheTheorem}, and thus $A(\rho)$ possesses a simple, real dominant eigenvalue $\mu$. Besides, it as left and right eigenvectors associated with $\mu$, respectively indicated as $\bm v$ and $\bm w$, which are strictly positive, that is, all their entries are $v_i > 0, w_i>0$. From this follows that the partial derivative of the dominant eigenvalue $\mu$ by the $i,j$-th element of $A$, $a_{ij} = [A]_{ij}$ is always positive:
\begin{align}
\label{eq:mu-derivative}
    \frac{\partial \mu}{\partial a_{ij}} = \frac{v_i w_j}{\bm v \bm w} > 0
\end{align}
where the left equality is according to \citep{Horn1985MatrixAnalysis} and is generally valid for simple eigenvalues. Here, $\bm v$ is assumed to be in row form, and $\bm v \bm w$ thus corresponds to a scalar product.

\subsection{Sufficient condition for dynamic homeostasis} \label{sec:dynamicHomeostasis}

In \citep{Greulich2021}, it was shown that a dynamic homeostatic state, where cell numbers may change over time but stay bounded, is assured if, \footnote{In \citep{Greulich2021}, this condition, defined through dependency on cell number, can be directly translated into a condition on the cell density derivative if the volume is assumed as a constant.}
\begin{equation} \label{eq:dynHomeostasis}
\mu'(\rho) < 0 \, \mbox{ for all } \rho > 0.
\end{equation} 
This sufficient condition requires that the dominant eigenvalue of $A$ as a function of the cell density, $\mu(\rho)$, is a strictly decreasing function of cell density. Also, the range of this function must be sufficiently large so that it has a root, i.e. a value $\rho^*$ with $\mu(\rho^*)=0$ must exist for the function $\mu(\rho)$.

Assuming that a non-trivial steady state, $\rho^* > 0$, exists, we now translate the sufficient condition for a dynamic homeostatic state, Eq. \eqref{eq:dynHomeostasis}, into conditions on the parameters as a function of the cell density, $\alpha_i(\rho)$. In particular, we can write,
\begin{align}\nonumber
\mu'(\rho) =&\sum_{ij} \frac{\partial \mu}{\partial a_{ij}} \frac{\partial a_{ij}}{\partial \rho} = \sum_{ij} \frac{v_i w_j}{\bm v \bm w} a'_{ij} = \sum_{i} \frac{v_i w_i}{\bm v \bm w} a'_{ii} + \sum_{i,j \neq i} \frac{v_i w_j}{\bm v \bm w} a'_{ij}  \\ \label{eqn:dmudrho_a}
\begin{split}
 = \sum_{i} \frac{v_i w_i}{\bm v \bm w} \left(\lambda'_i  -\gamma'_i - \sum_{j \neq i}\kappa'_{ij}\right)  + \sum_{i,j \neq i}\frac{v_j w_i}{\bm v \bm w}\kappa'_{ij} \,\, ,
\end{split}
\end{align} 
where we used Eq. \eqref{eq:mu-derivative} and the explicit forms of $a_{ij}$, the elements of the matrix $A$ according to Eq. \eqref{eq:MatrixElements}. Provided that all the parameters depend on $\rho$, condition \eqref{eq:dynHomeostasis} results in:
\begin{align}\label{eqn:dmudrho_alpha_wv_1}
0 > \mu' \implies 0 > \sum_{i} v_i w_i \left(\lambda'_i - \gamma'_i \right) + w_i \sum_{j\neq i} (v_j - v_i) \kappa'_{ij} \,\,\, \mbox{ for all $\rho > 0$} \, ,
\end{align}

While we cannot give an explicit general expression for the dominant eigenvectors $\bm v, \bm w$, this condition is sufficiently fulfilled if each term of the sum on the right-hand side of Eq. \eqref{eqn:dmudrho_alpha_wv_1} is negative. More restrictively, we have Eq. \eqref{eqn:dmudrho_alpha_wv_1} sufficiently fulfilled if
\begin{equation}
 \label{eqn:sufficientCond_noOmega}
 \begin{cases}
  \lambda'_i \leq 0, \,
    \gamma'_i \geq 0 \mbox{ for all } i  \\
  \lambda'_i < 0 \mbox{ or } \gamma'_i > 0 \mbox{ at for least one } i \\
  \kappa'_{ij} = 0\mbox{ for all } i, j 
 \end{cases} \qquad \mbox{ for } \rho>0
\end{equation}
This means that, excluding rates that are zero, which are biologically meaningless, if no direct state transitions within a cell type are subject to crowding feedback ($\kappa'_{ij}=0$), while all (non-zero) cell division rates depend negatively on $\rho$ ($\lambda'_{i}<0$), and differentiation rates depend positively ($\gamma'_{i}>0$), for all attainable levels of $\rho$, then dynamical homeostasis is ensured. 

Alternatively, we can rewrite Eq. \eqref{eqn:dmudrho_alpha_wv_1} as
\begin{align}\label{eqn:dmudrho_alpha_wv_2}
0 >  \sum_{i}\dfrac{v_i w_i}{\boldsymbol v \boldsymbol w} \left(\lambda'_i - \gamma'_i - \sum_{j\neq i} \kappa'_{ij} + \sum_{j\neq i} \dfrac{v_j}{v_i}\kappa'_{ij}\right) \,\,\, \mbox{ for all $\rho > 0$} \, ,
\end{align}
which, due to $\frac{v_j}{v_i} > 0$, implies another sufficient condition for dynamic homeostasis: 
\begin{equation} \label{eqn:sufficientCond_omegaNonZero} \begin{cases}
  \lambda'_i \leq 0, \,
    \gamma'_i \geq 0 \mbox{ for all } i  \\
  \lambda'_i < 0 \mbox{ or } \gamma'_i > 0 \mbox{ at for least one } i \\
  \kappa'_{ij} \leq 0 \mbox{ with } \lvert \sum_j  \kappa'_{ij} \rvert \leq \gamma'_i - \lambda'_i \mbox{ for all } i, j
 \end{cases}
\end{equation}
The above condition is less restrictive than Eq. \eqref{eqn:sufficientCond_noOmega}, allowing for some non-zero crowding feedback dependency of state transition rates $\kappa_{ij}$, as long as the crowding feedback strength of the total outgoing transition rate of each state does not outweigh the feedback on proliferation and differentiation rate of that state (if there is). 


\subsection{Necessary condition for strict homeostasis}

We now consider the circumstances under which a \emph{strict} homeostatic is maintained, that is, when a steady state of the cell population exists and is asymptotically stable.

A necessary condition for the existence of a steady state $\bm \rho^*$ (irrespective of stability) has been given in \citep{Greulich2021}, namely, that the cell type at the apex of the lineage hierarchy is self-renewing, i.e. its dynamical matrix $A$ has $\mu=0$. $\mu$ depends on the cell density $\rho$ of the apex cell type, since the dynamical parameters $\alpha_i$ and thus $A$ depend on $\rho$. As before, it is required that $\mu(\rho^*)$ has sufficient range so that a value $\rho^*$ with $\mu(\rho^*) = 0$ exists. This condition is fulfilled if the range of the feedback parameters $\alpha_i(\rho)$ is sufficiently large. In that case there exists an eigenvector $\boldsymbol \rho^{*}$ with $A(\rho^{*}) \boldsymbol \rho^{*} = 0$, which can be chosen by normalisation to fulfil $\sum_{i \in S} \rho^*_i = \rho^*$. Thus, $\bm \rho^*$ is a fixed point (steady state) of the cell population system \eqref{dyn_sys_feedback_eq}. Hence, we need to establish what is required for this state to be asymptotically stable.

To start with, we give the Jacobian matrix of the system \eqref{dyn_sys_feedback_eq} at the fixed point $\bm \rho^*$ :
\begin{align}\label{eqn:Jacobian_aij}
[J]_{ij} = \left.\frac{\partial [A(\rho)\boldsymbol \rho]_{i}}{\partial \rho_{j}}\right\rvert_{\bm \rho= \bm \rho^*} = a^{*}_{ij} + \eta_i \,\, ,
\end{align}
where 
\begin{align}
\label{eq:eta}
    \eta_i = \sum_{k} a'_{ik} \rho^{*}_{k} \,.
\end{align} 
Here and in the following, we assume the derivatives to be taken at the steady state, i.e. $a'_{ij} := \frac{da_{ij}}{d\rho}\vert_{\rho=\rho^*}$. The eigenvalues of the Jacobian matrix $J$ at $\rho^*$ determine the stability of the steady state $\rho^* $: it is asymptotically stable if and only if the real part of all eigenvalues of $J(\rho^*)$ is negative.

The Routh-Hurwitz theorem \citep{Franklin2014FeedbackSystems} states that for a polynomial to have only roots with negative real part, all its coefficients must necessarily be positive. Given that the eigenvalues of the Jacobian matrix $J$ are the roots of its characteristic polynomial, a necessary condition for $\rho^*$ to be asymptotically stable is that the coefficients of the characteristic polynomial of $J$ are all positive. 

Let us start by considering a self-renewing cell type with exactly two cell states being at the apex of a lineage hierarchy. This system has a 2 $\times$ 2 dynamical matrix $A$ and Jacobian $J$, whereby $A$ is irreducible and has dominant eigenvalue $\mu^A=0$. The characteristic polynomial of a generic 2$\times$2 matrix, $M$, is
\begin{equation} \label{eqn:SIcharacteristicPolyA2d}
    P^M(s) = s^2 + p_1^M s + p_0^M \,.
\end{equation}
with $p_1^M = -{\rm tr}(M)$ and $p_0^M = \det(M)$. In particular, since $A$ has an eigenvalue zero, 
\begin{equation} \label{eqn:SIpa22d_m}
    p^A_0 = \det(A) = a_{11}a_{22}- a_{12}a_{21} = 0 \,.
\end{equation}

From this follows that the right and left eigenvectors to the matrix $A$ associated with the dominant eigenvalue $\mu^A=0$, $\bm w$ and $\bm v$, are:
\begin{equation} \label{eqn:SIrighteigen2d_m}
\bm w = \begin{pmatrix} -a_{22} \\ a_{21}\end{pmatrix} \mbox{ and } \bm v = \begin{pmatrix} -a_{22} & a_{12}\end{pmatrix} \, .
\end{equation} 

From the Jacobian matrix $J$, we get equivalently,
\begin{align} \label{eqn:SIpj22d_m}
    p^J_0 &= \det(J) = \dfrac{(a_{21} - a_{22})(- a_{22}\eta_1 + a_{12}\eta_2)}{a_{22}} = \bm v \boldsymbol \eta\dfrac{\lvert \bm w \rvert}{a_{22}} \, ,
\end{align}
with the $L^1$-norm $\lvert \bm w \rvert = w_1 + w_2 = -a_{22} + a_{21}$\footnote{Note that $a_{ii}$ is always negative or zero}. Here we used the form of $J$ in Eq. \eqref{eqn:Jacobian_aij} with $\bm \eta = (\eta_1,\eta_2)$ from \eqref{eq:eta}, as well as the relations \eqref{eqn:SIpa22d_m} and \eqref{eqn:SIrighteigen2d_m}, and we factorised the determinant.

From Eq. \eqref{eqn:dmudrho_a}, we can further establish: 
\begin{align} \label{eqn:mudot}
\mu' &= \sum_{ij} \dfrac{v_i w_j}{\boldsymbol v \boldsymbol w} a_{ij}' = \sum_{ij} \frac{\lvert\bm w \rvert}{\rho^*}\dfrac{v_i \rho^*_j}{\boldsymbol v \boldsymbol w} a_{ij}' = \frac{\lvert\bm w \rvert}{\rho^*}\frac{\bm v \bm \eta}{\bm v \bm w} \\
& = -\frac{a_{22} p_0^J}{\rho^* p_1^J a_{22}} \,.
\end{align}
Here, we used that $\bm \rho^*$ is a dominant right eigenvector, and thus $\bm \rho^* = \frac{\rho^*}{\lvert w \rvert} \bm w$, and furthermore we used the definition of $\eta_i = \sum_j a'_{ij} \rho_j^*$, we substituted Eq. \eqref{eqn:SIpj22d_m}, and used that $\bm v \bm w = a^2_{22} + a_{12}a_{21} = - p_1^A a_{22}$. Finally, we get:
\begin{align}\label{eqn:c0_mudot1}
p^J_0 = -\mu' \rho^* p^A_1 \,.
\end{align}
Notably, we can show that this relation also holds for higher dimensions by explicitly computing the coefficients of characteristic polynomials $p_i^{A,J}$, the eigenvalues and eigenvectors, and then evaluating both sides of the equation. For systems with three states, this can be done analytically. For systems with 4,5, and 6 states we tested relation \eqref{eqn:c0_mudot1} numerically by generating $N=$1000 random matrices with entries chosen from a uniform distribution\footnote{The diagonal elements of the random matrix are tuned using a local optimiser (\textit{fmincon} function of Matab) so that the matrix has a zero dominant eigenvalue.}. In each case, this relation was fulfilled. Hence we are confident that this relation holds up to 6 states, and it is reasonable to expect this to hold also for larger systems.

Since $A$ has a simple dominant eigenvalue $\mu^A=0$, we can factorise one term from the characteristic polynomial, $P(s)=s Q(s)$ knowing that all roots of $Q(s)$ are negative. Applying the Routh-Hurwitz necessary condition to $Q(s)$, it follows that the coefficients of the polynomial $Q$, $p_i^Q > 0$, where $i = 1, 2, ..., n-1$. Thus, $p_{1}^A > 0$ and considering that $\rho^* > 0$ by definition, then for having $p^J_0 > 0$ we must require $\mu' < 0$. Therefore, a necessary condition for a stable, strict homeostatic state is
\begin{align}\label{eqn:necessaryCond}
    0 > \mu' \implies 0 > \left. \sum_{i} v_i w_i \left(\lambda'_i - \gamma'_i \right) + w_i\sum_{j\neq i} (v_j - v_i) \kappa'_{ij}\right\vert_{\rho=\rho^*} ,
\end{align}
where on the right-hand side, we used Eq. \eqref{eqn:dmudrho_alpha_wv_1}. This condition is bound to the validity of Eq. \eqref{eqn:c0_mudot1}, that is, we can show it analytically for up to three states and numerically up to 6 states. Nonetheless, we also expect this to be true for larger systems.

One way to satisfy this necessary condition is if at $\rho=\rho^*$
\begin{equation} \label{eqn:necessaryCond_example} \begin{cases}
  \lambda'_i \leq 0, \,
    \gamma'_i \geq 0 \mbox{ for all } i  \\
  \lambda'_i < 0 \mbox{ or } \gamma'_i > 0 \mbox{ at for least one } i \\
  \kappa'_{ij} = 0 
 \end{cases} \,.
\end{equation}

Notably, the necessary conditions \eqref{eqn:necessaryCond} and \eqref{eqn:necessaryCond_example} only differ from the sufficient conditions for dynamic heterogeneity, Eqs. \eqref{eqn:dmudrho_alpha_wv_1} and \eqref{eqn:sufficientCond_noOmega}, by needing to be fulfilled \emph{only at} the steady-state cell density $\rho^*$, whereas to ensure dynamic homeostasis, those should be valid for a sufficiently large range of $\rho$.

\subsection{Sufficient condition for strict homeostasis}\label{sec:asymptoticStability}

Now we assess under which circumstances a strict homeostatic state is assured to prevail. 

First of all, the necessary conditions from above need to be fulfilled. In particular, the parameter functions $\alpha_i(\rho)$ must have a sufficient range so that $\mu(\rho)$ has a root, i.e. $\rho^*$ with $\mu(\rho^*)=0$ exists, from which the existence of a steady state follows. The question now is whether we can find sufficient conditions assuring that the fixed point $\boldsymbol \rho^{*}$ with $\sum_i \rho^*_i = \rho^*$ is (asymptotically) stable.

Let us define a matrix $B(\bm x)$, $\bm x = (x_1,...,x_m)$ with $b_{ij}(\bm x)=[B]_{ij}(\bm x) = a^*_{ij} + x_i$. Hence, $B(x_i = 0) = A(\rho^*)$ and $B(x_i = \eta_i) = J$, where $J$, the Jacobian matrix, and $\eta_i$ are defined as in \eqref{eqn:Jacobian_aij} and \eqref{eq:eta}, respectively. We consider now the dominant eigenvalue as function of the entries of $B$, $\mu[B] := \mu(\lbrace b_{ij}\rbrace\vert_{i,j=1,...,m})$ (the square brackets are chosen to denote the difference from the function $\mu(\rho)$). For sufficiently small $\eta_i$, we can then express the dominant eigenvalue of the Jacobian matrix $J$, $\mu[J]$, relative to the dominant eigenvalues of $A^* := A(\rho^*)$ as, 
\begin{align}
    \mu[J] = \mu[A^*] + \sum_i \frac{\partial \mu}{\partial x_i}\rvert_{x_i = 0} \, \eta_i + O(\bm \eta^2) \,\, ,
\end{align}
with,
\begin{align}
 \frac{\partial \mu}{\partial x_i}\rvert_{x_i = 0} = \sum_{ij} \frac{\partial \mu}{\partial b_{ij}}\frac{\partial  b_{ij}}{\partial x_i}\rvert_{x_i = 0} = \sum_{ij} \frac{\partial \mu}{\partial a_{ij}}\vert_{B=A^*} \,\, ,
\end{align}
since for $\bm x= \bm 0$, $b_{ij}=a_{ij}$ for all $i,j$. It follows that for sufficiently small\footnote{That is, there exist $\epsilon_i > 0$ so that this is valid for any $\vert \eta_i \vert < \epsilon_i$} $\eta_i$, and if all $\eta_i < 0$, we have
\begin{align}
\label{eq:muJ_2}
    \mu_J = \mu[A^*](\rho^*) + \sum_i \frac{\partial \mu_B}{\partial x_i}\vert_{x_i = 0} \eta_i + O(\eta_i^2) \approx \sum_i \frac{\partial \mu_A}{\partial a_{ij}} \eta_i < 0
\end{align}
since all $\frac{\partial \mu_A}{\partial a_{ij}} > 0$ (according to Eq. \eqref{eq:mu-derivative}) and $\mu_A(\rho^*)=0$. Hence, since $\mu_J < 0$, the steady state $\bm \rho^*$ is asymptotically stable if all $\eta_i < 0$. Thus, we get a sufficient condition for asymptotic stability of the steady state $\rho^{*}$:
\begin{align}
\label{eq:asymp-stab}
0 > \eta_i = \rho^{*}_{i} (\lambda'_{i} - \gamma'_{i} ) + \sum_{k \neq i} (\kappa'_{ki} \rho^{*}_{k}-\kappa'_{ik}\rho^{*}_{i}) > -\epsilon_i \mbox{ for all } i
\end{align}
where $\epsilon_i > 0$ is sufficiently small. As this is an asymptotically stable steady state, it corresponds to a controlled strict homeostatic state. In this case, even if the cell numbers are disturbed (to some degree), the cell population is regulated to return to the strict homeostatic state. 

Notably, condition \eqref{eq:asymp-stab} is fulfilled if,
\begin{equation} \label{eqn:asymp-stab_2} \begin{cases}
  \lambda'_i \leq 0, \,
    \gamma'_i \geq 0 \mbox{ for all } i  \\
  \lambda'_i < 0 \mbox{ or } \gamma'_i > 0 \mbox{ at for least one } i \\
  \kappa'_{ij} = 0 \\
  \mbox{ and } \vert \lambda'_i \vert, \vert \gamma'_i \vert, < \epsilon_i
 \end{cases}
\end{equation}
Furthermore, we may soften the condition on $\kappa_{ij}$ to $\frac{\kappa'_{ij}}{\kappa'_{ji}} < \frac{\rho^*_j}{\rho^*_i}$ to allow also some degree of feedback for the $\kappa_{ij}$.

The conditions \eqref{eqn:asymp-stab_2} are very similar to the ones for dynamic homeostasis, \eqref{eqn:sufficientCond_noOmega}, but here these conditions only need to be fulfilled at $\rho=\rho^*$, whereas for dynamic homeostasis they need to be fulfilled for a sufficient range of $\rho$. Moreover, in addition to the qualitative nature of the feedback (related to the signs of $\lambda_i',\gamma_i'$), the `strength' of the crowding feedback, i.e. the absolute values of $\lambda_i',\gamma_i'$ must not be `too large', that is, smaller than $\epsilon_i$. We cannot, in general and for all system sizes, give a definite value for the feedback strength bound $\epsilon_i$ below which strict homeostasis is assured. Nevertheless, by using the sufficient stability criterion based on the Routh-Hurwitz criterion \citep{Franklin2014FeedbackSystems} we can identify those bounds for systems with up to three cell states, which guides expectations for larger systems. The details of this criterion and the necessary derivations are shown in \aref{appendix:RouthHurwitz}. There, we show that for systems with one or two cell states, $\epsilon_i=\infty$, which means that asymptotic stability is ensured for arbitrary feedback strengths. For systems with three cell states, we can assure that $\epsilon_i=\infty$ if certain further conditions are met (see Eq. \eqref{eq:sufficient_cond_eps_inf}). Otherwise, $\epsilon_i$ can be determined implicitly from the roots of a quadratic form (Eq. \eqref{eqn:quadraticEqnEpsilon}), and thus stability may depend on the magnitude of the feedback. In principle, such bounds can be found for larger systems too, but the algebraic complexity of this process renders it unfeasible to do this in practical terms.

\subsection{Robustness to perturbations and failures}\label{sec:robustness}

Now, we wish to assess the \emph{robustness} of the above crowding control mechanism, i.e. what occurs if it is disrupted, for example, by the action of toxins, other environmental cues, or by cell mutations. 
More precisely, we will study what happens if one or more feedback pathways, here characterised as a parameter $\alpha_i$ with $\alpha_i' \neq 0$ fulfilling the conditions for (dynamic or strict) homeostatic control, is failing, that is, it becomes $\alpha'_i = 0$. We will first address the case of tissue-extrinsic factors, i.e. those affecting all the cells in the tissue, and then the case of single-cell mutations. In the latter case, only a single cell would initially show a dysregulated behaviour, yet, if this confers a proliferative advantage, it can lead to hyperplasia and possibly cancer \citep{Tomasetti2013,Colom2016,Rodilla2018CellularCancer}. 

First, we note that the sufficient condition for strict homeostasis, given by Eq. \eqref{eqn:asymp-stab_2}, is overly restrictive. 
In a tissue cell type under crowding feedback control with $\lambda'_i < 0$ and $\gamma'_i > 0$ for more than one $i$, there is a degree of redundancy. That is, if the feedback is removed for one or more of these parameters (changing to $\lambda'_i = 0$ and, or $\gamma'_i = 0$), then the sufficient condition for a strict homeostatic state remains fulfilled as long as at least one $\lambda'_i$ or $\gamma'_i$ remains non-zero. This possible redundancy confers a degree of robustness, meaning that feedback pathways can be removed -- setting $\alpha'_i = 0$ -- without losing homeostatic control. Since the necessary conditions, Eqs. \eqref{eqn:necessaryCond}, are even less restrictive, tissue homeostasis may even tolerate more severe disruptions that reverse some feedback pathways, e.g. switching from $\lambda'_i < 0$ to $\lambda'_i > 0$, as long as other terms in the sum on the right-hand side of \eqref{eqn:necessaryCond} compensate for this changed sign, ensuring that the sum as a whole is negative. In any case, it is important to remind the underlying assumption for which a non-trivial steady state exists. In case the variability of the kinetic parameters is not enough to assure the condition $\mu(\rho^* = 0)$, then the tissue will degenerate, either shrinking and eventually disappearing or indefinitely growing.

From the above considerations, we conclude that if crowding control applies to more than one parameter $\alpha_i$, that is, $\alpha'_i \neq 0$ with appropriate sign and magnitude, homeostasis is potentially robust to feedback disruption. This may include a simple variation of the feedback function $\alpha'_i$ but also perturbation in the feedback functions shape and complete feedback failure, $\alpha'_i = 0$.

\begin{figure}[h]
\begin{center}
\includegraphics[width=0.45\columnwidth]{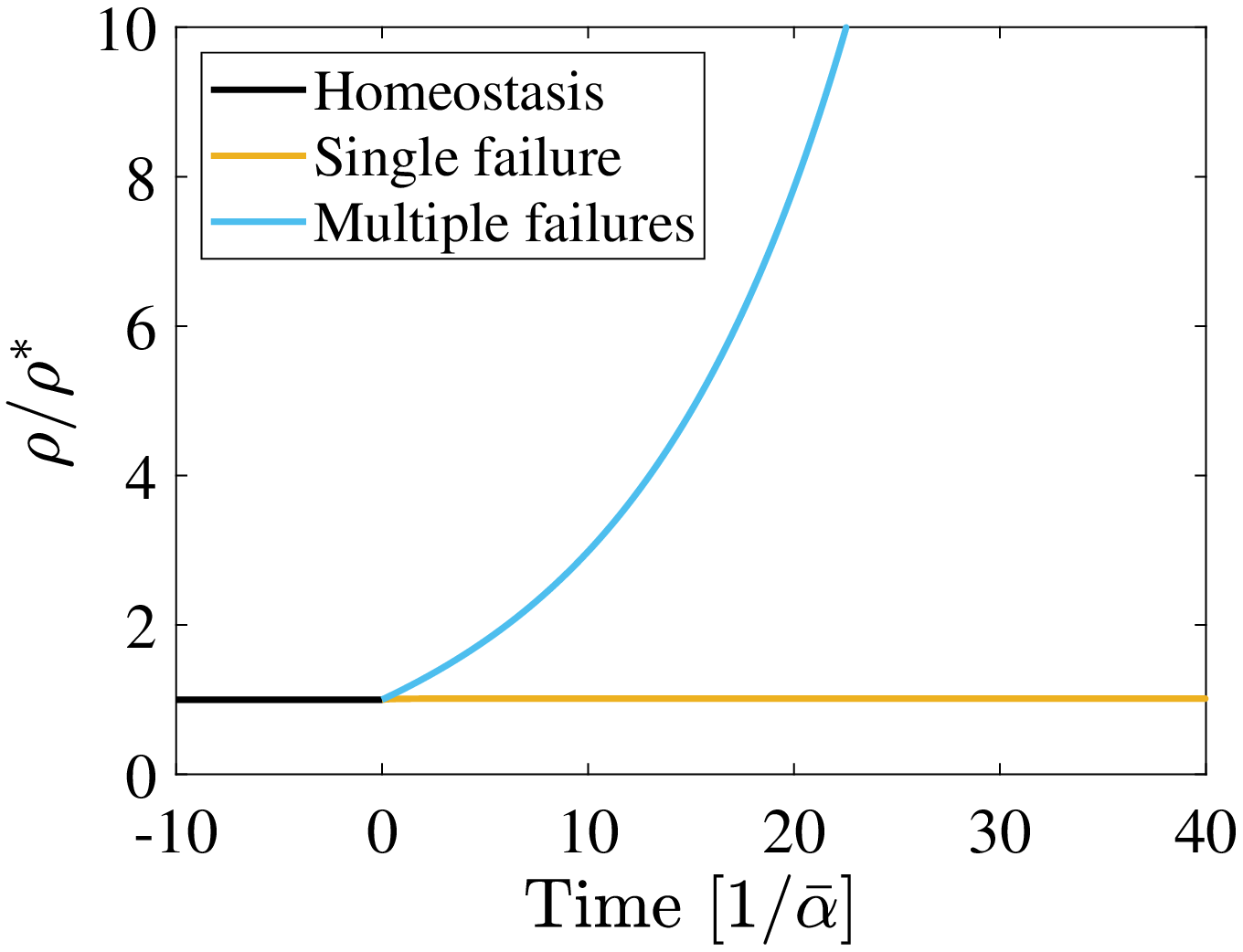}
\includegraphics[width=0.45\columnwidth]{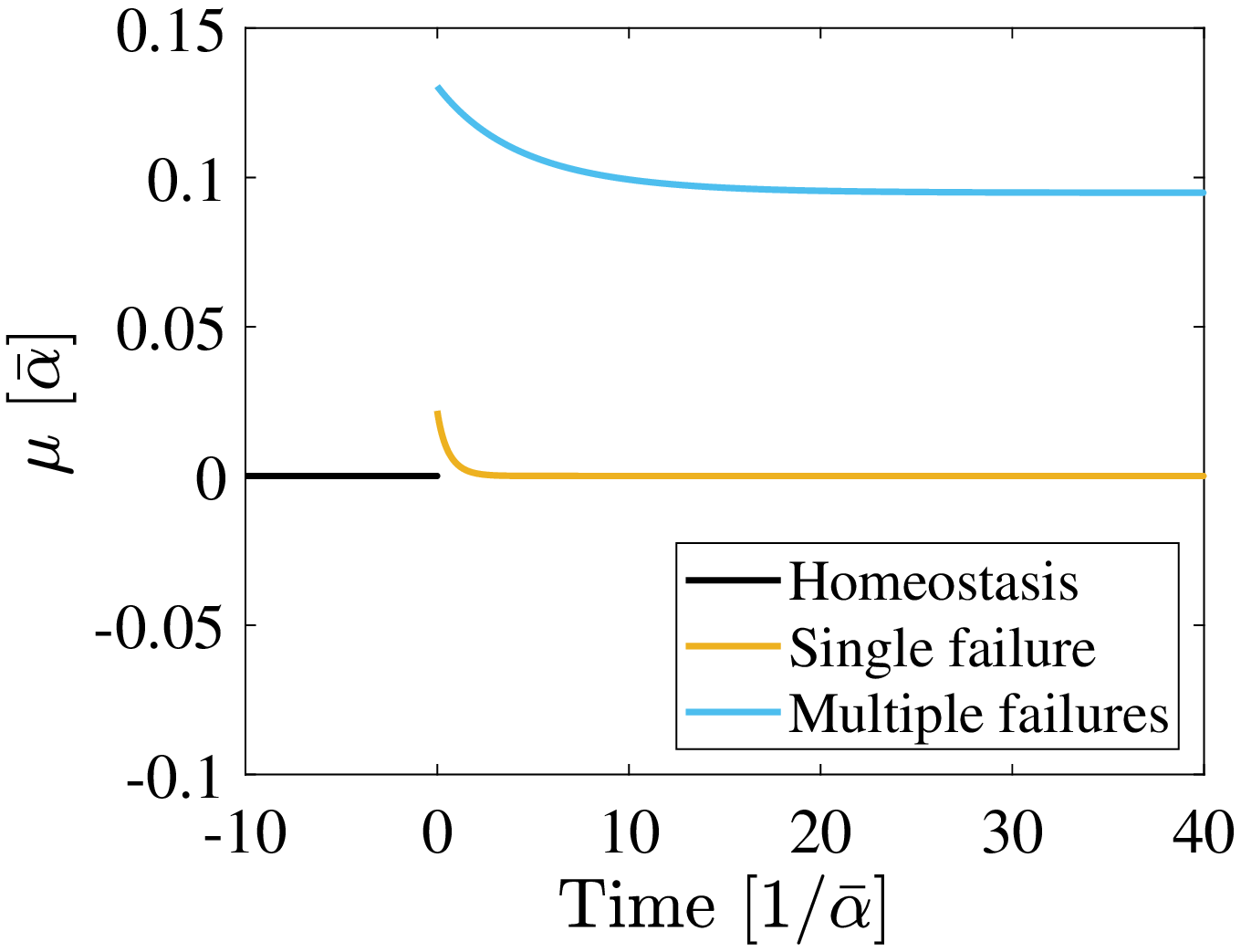}
\caption{\label{fig:FeedbackFailureTissue}
Cell dynamics in terms of cell density, scaled by the steady state in the homeostatic case, as a function of time (left) and the corresponding variation of the dominant eigenvalue $\mu$ (right). Time is scaled by the inverse of $\bar{\alpha} = \min_i{\alpha_i^*}$. The homeostatic model is perturbed at a time equal to zero to include feedback failure of the type $\alpha'_i = 0$. In the case where only one feedback function fails (``Single failure''), the system is able to achieve and maintain a new homeostatic state, characterised by a constant cell density and a zero dominant eigenvalue. In case more than one feedback fails (``Multiple failures''), the cell dynamics are unstable since a steady state does not exist and $\mu > 0$ for all $\rho$. The cell fate model corresponds to model \eqref{mdl:ExampleModel} with parameters given in \tref{tab:FeebackTestCases_table} and \tref{tab:FeebackTestCases_ff_table}.
}
\end{center}
\end{figure}

An illustrative example of this situation is shown in \fref{fig:FeedbackFailureTissue}. Here, the time evolution of the cell density is shown for a three-state cell fate model, which has been computed by integration of the dynamical system \eqref{dyn_sys_feedback_eq} (the details of this model are given in \aref{appendix:IllustrativeCase} as Eq. \eqref{mdl:ExampleModel} and illustrated in \fref{fig:FeebackTestCases_figNetwork}). Four kinetic parameters are regulated via crowding control satisfying the sufficient condition for strict homeostasis, \eqref{eqn:asymp-stab_2}. Then, starting from this homeostatic configuration, feedback disruption is introduced at a time equal to zero. In one case (``Single failure''), a single kinetic parameter suffers a complete failure of the type $\alpha_i' = 0$. In this case, the remaining feedback functions compensate for this failure, and a new homeostatic condition is achieved. Instead, in the second case (``Multiple failures''), failures are applied so that three of the four kinetic parameters initially regulated do not adjust with cell density\footnote{Only in this example, feedback control fails upon multiple failures, while in general, multiple failures may still be compensated to maintain homeostatic control.}. Notably, the only feedback function left satisfies the condition for asymptotic stability, \eqref{eqn:asymp-stab_2}. Nevertheless, the variability of this kinetic parameter is not enough to assure the existence of a steady state, since in this case, the function $\mu(\rho)$ does not possess any root. Hence $\mu > 0$ for all $\rho$, leading to an indefinite growth of the cell population. Additional test cases are presented in \aref{appendix:FeedbackFailure}.

So far, we modelled the feedback dysregulation as acting on a global scale, thus changing the whole tissue's dynamics behaviour. This situation represents a feedback mechanism affected by cell-extrinsic signals, in which any dysregulation applies to all the cells in the same way. However, dysregulation can also act at the single-cell level, for example, when DNA mutations occur. In this case, the impact of the dysregulation is slightly different, as explained in the following. 

Suppose, upon disruption of crowding control in a single cell, for example, by DNA mutations, a sufficient number of crowding feedback pathways remain so that there is a steady state and the sufficient condition \eqref{eqn:asymp-stab_2} is still fulfilled. In that case, homeostasis is retained, just as when this occurs in a tissue-wide disruption. However, if the homeostatic control of that single cell fails such that the cell becomes hyperproliferative, $\mu > 0$, or declining, $\mu < 0$, the tissue may still remain homeostatic. If $\mu < 0$, the single mutated cell will be lost, upon which only a population of crowding controlled cells remain, which remain in homeostasis. If $\mu > 0$ in a single cell, hyper-proliferation is not ensured either: while the probability for mutated cells to grow in numbers is larger than to decline, due to the low numbers, mere randomness can lead to the loss of the mutated cell with a non-zero probability, which results in the extinction of the dysregulated mutant\footnote{For example, in the case of a single state with cell division rate $\lambda$ and loss rate $\gamma$ -- a simple branching process -- the probability for a mutant with $\mu > 0$, that is, $\lambda > \gamma$, to establish is $1-\gamma/\lambda$, which is less than certainty.}. In that case, the mutant cells go extinct and the tissue remains homeostatic despite the disruption of homeostatic control in the mutated cells; a stark contrast to disruption on the tissue level. Otherwise, if the mutant clone (randomly) survives, it will continue to hyper-proliferate and eventually dominate the tissue, which is thus rendered non-homeostatic. However, the tissue divergence time scale may be much longer than the case where the same dysregulation occurs in all cells. 

The deterministic cell population model \eqref{dyn_sys_feedback_eq} is suitable for describing the average cell numbers. Nevertheless, it fails to describe the stochastic nature of single-cell fate choice. Thus, assessing a single cell's impact on tissue dynamics requires stochastic modelling. To that end, we implemented this situation as a Markov process with the same rates as the tissue cell population dynamics model\footnote{While a Markov process is an approximation which not necessarily reflects the probability distribution of subsequent event times realistically, it is often sufficient to assess the qualitative behaviour of a system with low numbers, subject to random influences from the environment.} (see \aref{appendix:singleCellMutation} for more details). 

\begin{figure}[h]
\begin{center}
\includegraphics[width=0.45\columnwidth]{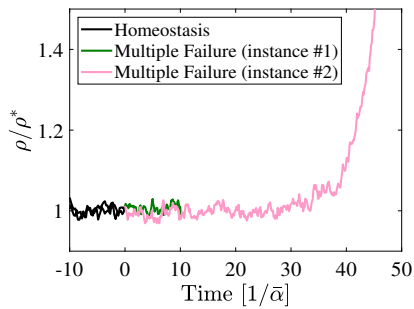} 
\caption{\label{fig:SingleCellMutation} Numerical simulation results of a stochastic version of the model used in \fref{fig:FeedbackFailureTissue} upon disruption of crowding control in a single cell, mimicking a DNA mutation. At a time equal to 0, the initially homeostatic model is disrupted with a single cell presenting multiple failures in the feedback control, as in \fref{fig:FeedbackFailureTissue}. Two instances of simulations run with identical parameters are presented. The rescaled cell density $\rho/\rho^*$ is shown as a function of the time, scaled by the inverse of $\bar{\alpha} = \min_i{\alpha_i^*}$. Whilst the mutated cell and its progeny go extinct in one instance ($\#$1), in the other ($\#$2), mutated cells prevail and hyper-proliferate so that tissue homeostasis is lost. The simulation stops when the clone goes extinct or when instability is detected. Full details of the simulation are provided in \aref{appendix:singleCellMutation}.
}
\end{center}
\end{figure}

In \fref{fig:SingleCellMutation}, we show numerical simulation results of a stochastic version of the model used for previous results in \fref{fig:FeedbackFailureTissue}, depicted in terms of tissue cell density as a function of time. Here, two possible realisations of the same stochastic process are presented. We note that the initially homeostatic tissue results in stochastic fluctuations of the cell density, which remain, on average, constant. At a time equal to zero, a single cell in this tissue switches behaviour, presenting multiple failures which, if applied to all the cells, would determine the growth of the tissue (corresponding to Multiple failures curve in \fref{fig:FeedbackFailureTissue}). In one instance of the stochastic simulation, however, the mutated clone goes extinct after some time, leaving a tissue globally unaffected by the mutation. On the other hand, in another instance, the mutated clone prevails, leading to the growth of the tissue cell population. The fact that vastly different outcomes can occur with the same parameters and starting conditions demonstrates the impact of stochasticity in the case of a single-cell mutation.

\subsection{Quasi-dedifferentiation}

In the previous section, we addressed the case where external or cell-intrinsic factors disrupt homeostatic control in self-renewing cells of a tissue. However, situations such as injury, poisoning, or cell radiation might also affect homeostasis in other ways. An example is when stem cells are completely depleted from the tissue. In this context, many studies about tissue regeneration after injury report evidence of cell plasticity \citep{Tetteh2015,Tetteh2016}, when committed cells regain the potential of the previously depleted stem cells. Cell \emph{dedifferentiation} is just an example where differentiated cells return to an undifferentiated state as a response to tissue damage. Lineage tracing experiments confirmed this feature {\it in vivo} in several cases \citep{Tata2016CellularDay, Merrell2016AdultStyle, Tata2013DedifferentiationVivo, Puri2015PlasticityDisease}.

\begin{figure}
\centering
\includegraphics[width=0.4\textwidth]{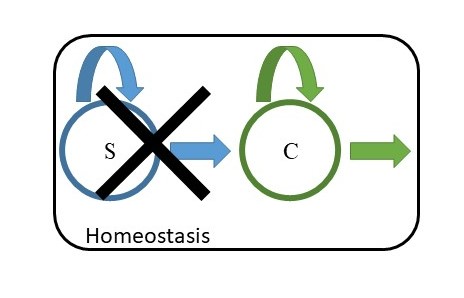}
\caption{Sketch representative of the quasi-dedifferentiation scenario. A homeostatic system enclosed in the black box comprises two cell types: a stem cell type, $S$, (blue) and a committed cell type, $C$, (green). In the unperturbed homeostatic scenario, $S$ is self-renewing, characterised by a growth parameter at the steady state $\mu^* = 0$, and $C$ is transient, with a growth parameter at the steady state $\mu^* < 0$. Both cell types are subject to crowding control, fulfilling both conditions \eqref{eqn:sufficientCond_noOmega}, and \eqref{eqn:asymp-stab_2}. By removing the stem cell type $X_S$, the committed cell type acquires self-renewing property through crowding control, effectively becoming a stem cell type (see Figure \ref{fig:dedifferentiation}).} \label{fig:dediff_sketch}
\end{figure}

In the following, we assess how committed progenitor cells respond to the depletion of the stem cell pool if they are under crowding feedback control. Without loss of generality, let us consider an initially homeostatic scenario where there is a self-renewing (i.e. stem) cell type (S) -- with growth parameter $\mu=0$ -- at the apex of a lineage hierarchy, and a committed progenitor cell type (C) -- with $\mu<0$, but with at least one state that has a non-zero cell division rate -- below type S in the hierarchy, as depicted in \fref{fig:dediff_sketch}. Based on this cell fate model, S-cells proliferate and differentiate into C-cells while maintaining the S-cell population. The C-cells also proliferate and differentiate into other downstream cell types which we do not explicitly consider here. C-cells do not maintain their own population; only the steady influx of new cells of that type via differentiation of S-cells into C-cells maintains the latter population (see \citep{Greulich2021}). We further assume that both S- and C-cells are under appropriate crowding control, fulfilling both the sufficient conditions for dynamic homeostasis, \eqref{eqn:sufficientCond_noOmega}, and for stable, strict homeostasis, \eqref{eqn:asymp-stab_2}.

Based on the above modelling, we can write the dynamics of the cell densities belonging to the committed progenitor type as,
\begin{align}
\label{eqn:dyn_feedback_cellTypeU}
\frac{d}{dt}\bm \rho_c = A_c(\rho_c) \bm \rho_c +  \bm u \, ,
\end{align}
where 
$\bm \rho_c = (\rho_{m_s+1},\rho_{m_s+2},..,\rho_{m_s+m_c})$ are the cell densities in the committed C-type, with ${m_s}$ being the number of states of the self-renewing S-type. $A_c$ is the dynamical matrix restricted to states in the C-type and $u_i = \sum_{j=1}^{m_s}\kappa_{ji} \rho_j$ is a constant vector quantifying the influx of cells into the C-type.

First, we note that the Jacobian matrix of a committed cell type, described by \eqref{eqn:dyn_feedback_cellTypeU}, $J=\left[\frac{\partial A(\rho_c) \bm \rho_c}{\partial \rho_j}\right]_{j=m_s + 1,...,m_s+m_c}$, has the same form as a cell type at the apex of the hierarchy, since $\bm u$ does not depend on the densities $\rho_{m_s+1,...,m_s+m_c}$. From this follows that if C-cells are regulated by crowding control, fulfilling the conditions \eqref{eqn:asymp-stab_2}, then also the population of C-cells is stable around a steady state $\rho_c^*$, albeit with a growth parameter $\mu_c(\rho_c^*) < 0$\footnote{This can be seen when multiplying the steady state condition for \eqref{eqn:dyn_feedback_cellTypeU}, $A_c(\rho_s,\rho_c) \bm \rho_c +  \bm u = 0$ with a positive left dominant eigenvector $\bm v$, giving, $\mu_c \bm v \bm \rho^*_c +  \bm v \bm u = 0$. Since $\bm \rho^*$ and $\bm v$ have all positive entries and $\bm u$ is non-negative, this equation can only be fulfilled for $\mu_c<0$.}.

We now consider the scenario where all stem cells are depleted at some point, as was experimentally done in \citep{Tata2013DedifferentiationVivo,Tetteh2016}. This would stop any replenishment of C-cells through differentiation of S-cells, corresponding to setting $\bm u=0$ in \eqref{eqn:dyn_feedback_cellTypeU}. Hence we end up with the dynamics $\dot{\bm \rho_c} = A(\rho_c) \bm \rho_c$. 
Now, assuming that the function $\mu(\rho)$ has sufficient range, so that $\mu(\rho_c^{**})=0$ for some $\rho_c^{**}$, and provided that $A(\rho_c)$ is under crowding control fulfilling the sufficient conditions for asymptotic stability of a steady state, then, following our arguments from section \ref{sec:asymptoticStability}, the population of C-cells will attain a stable steady state. In other words, those previously committed cells become self-renewing cells. Also, since they now reside at the apex of the lineage hierarchy (given that S-cells are absent), they effectively become stem cells. 

Hence, under crowding control, previously committed progenitor cells (committed cells that can divide) will automatically become stem cells if the original stem cells are depleted. Commonly, such a reversion of a committed cell type to a stem cell type would be called `dedifferentiation' or `reprogramming'. However, in this case, no genuine reversion of cell states occurs; previously committed cells do not transition back to states associated with the stem cell type. Instead, they respond by crowding feedback and adjust their dynamical rates so that $\mu$ becomes zero, hence attaining a self-renewing cell type. Crucially, this new stem cell type is fundamentally different to the original one and still most similar to the original committed type. We call this process \emph{quasi-dedifferentiation}. The quasi-dedifferentiation follows the same reversion of proliferative potential as in `genuine' dedifferentiation but without explicit reversion in the cell state trajectories.

\begin{figure}
\centering
\includegraphics[width=0.45\textwidth]{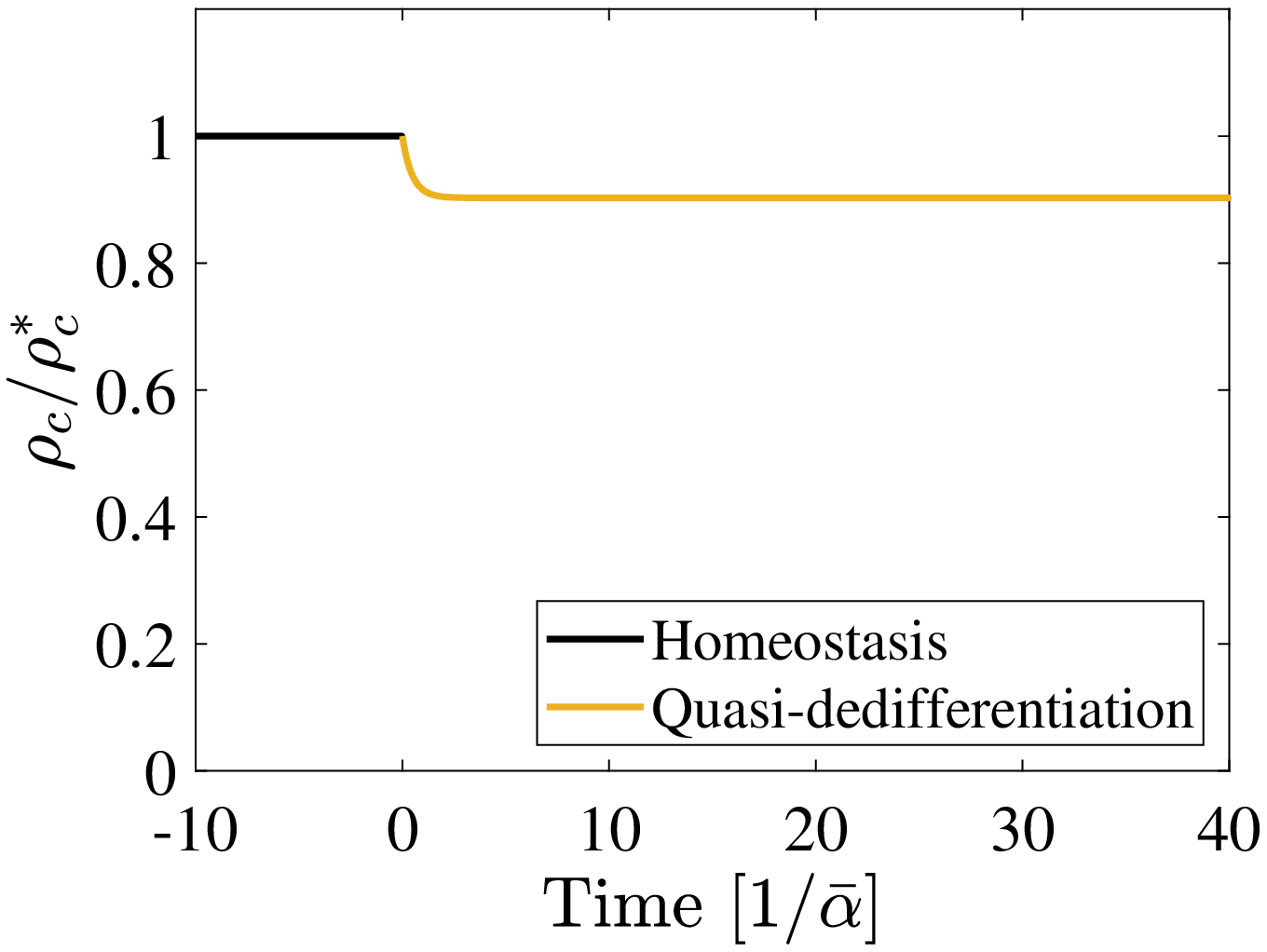}
\includegraphics[width=0.45\textwidth]{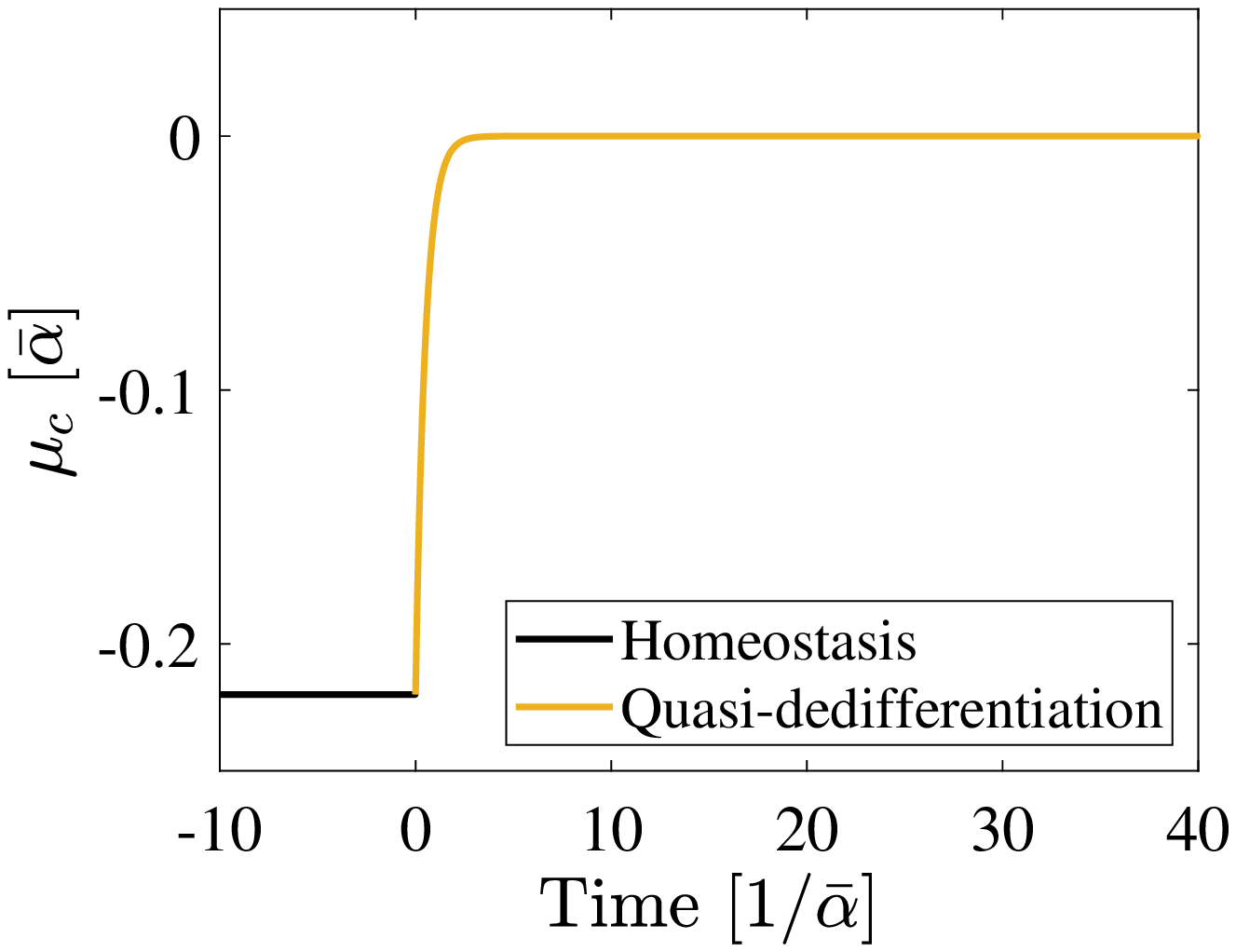}
\caption{Cell dynamics of an initially committed cell type C ($\mu<0$) upon removal of all stem cells. (Left) Cell density scaled by the steady-state density as a function of time. (Right) Corresponding variation of the dominant eigenvalue $\mu_c$. Time is scaled by the inverse of $\bar{\alpha} = \min_i{\alpha_i^*}$. It is assumed that a stem cell type, S, initially resides in the lineage hierarchy above the committed cell type (as in \fref{fig:dediff_sketch}). S cells differentiate into C cells, which is modelled as a constant cell influx of C-cells (S is not explicitly simulated). At a time equal to zero, a sudden depletion of S cells is modelled by stopping the cell influx. After some transitory phase, the cell population stabilises around a new steady state and becomes self-renewing with $\mu_c = 0$. The full description of the dynamical model, which corresponds to model \eqref{mdl:ExampleModel} with parameters given in \tref{tab:FeebackTestCases_table}, is reported in \aref{appendix:qDedifferentiation}.} \label{fig:dedifferentiation}
\end{figure}

The following numerical example illustrates this situation. We focus on the cell dynamics of a single C-type regulated via crowding feedback (detail of the model are provided in \aref{appendix:qDedifferentiation}). The cell density as a function of the time, shown in \fref{fig:dedifferentiation}, is obtained by integrating the corresponding cell population model according to Eq. \eqref{dyn_sys_feedback_eq}. The system is initially in a homeostatic condition, meaning that there is a constant influx of cells from some upstream self-renewing types. Such upstream types are assumed to be properly regulated such that this cell influx is constant over time. At a time equal to zero, the cell influx becomes suddenly zero, representing an instantaneous removal of all the self-renewing cells from the tissue. A new homeostatic condition is achieved after a transitory phase thanks to the crowding feedback acting on the C-type. This example demonstrates how an initially committed cell type, i.e. with $\mu_c < 0$, regulated via crowding feedback, might be able to switch, upon disruption, to a self-renewing behaviour $\mu_c = 0$.

\section{Discussion}\label{sec:Discussion}

For maintaining healthy adult tissue, the tissue cell population must be maintained in a homeostatic state. Here, we assessed one of the most common generalised regulation mechanisms of homeostasis, which we refer to as \emph{crowding feedback}. Based on this, progenitor cells (stem cells and committed progenitors) adjust their propensities to divide, differentiate, and die, according to the surrounding density of cells, which they sense via biochemical or mechanical signals. For this purpose, we used a generic mathematical model introduced before in Refs. \citep{Parigini2020UniversalityStrategies,Greulich2021}, which describes tissue cell population dynamics in the most generic way, including cell divisions, cell state transitions, and cell loss / differentiation. Based on this model, we rigorously define what is meant when speaking of a `homeostatic state', introducing two notions: a \emph{strict} homeostasis is a steady state of the tissue cell population dynamics, while \emph{dynamical} homeostasis allows, in addition to strict homeostasis, for oscillations and fluctuations, as long as a finite long-term average cell population is maintained (such as the endometrium during the menstrual cycle). 

By analysing this dynamical system, we find several sufficient and necessary conditions for homeostasis. These conditions are formulated in terms of how the propensities of cell division, differentiation, and cell state changes, of cells whose type is at the apex of an adult cell lineage hierarchy, may depend on their cell density. We find that when, for a wide range of cell density values, the cell division propensity of at least one state decreases with cell density or the differentiation propensity increases with it, while other propensities (e.g. of cell state transitions) are not affected by the cell density, then dynamic homeostasis prevails \eqref{eqn:sufficientCond_noOmega}. For strict homeostasis to prevail, this only needs to be fulfilled at the steady state itself, but in addition, the magnitude of the feedback strength may not be too large \eqref{eqn:asymp-stab_2}. We can derive explicit and implicit expressions for the bound on feedback strength for systems of two and three-cell states but cannot do so for arbitrary systems. Furthermore, we find that a necessary condition for strict homeostasis is that the conditions for dynamic homeostasis are met at least at the steady state cell density.

A direct consequence of the conditions we found is that they allow for a considerable degree of redundancy when more than one propensity depends appropriately on the cell density. Hence feedback pathways, that is, cell dynamics parameters depending on the cell density, may serve as `back-ups' to each other if one fails. We demonstrate that this confers robustness to the homeostatic system in that one or more crowding feedback pathways may fail, yet the tissue remains in homeostasis.

Finally, we assess how crowding feedback regulation affects the response of committed progenitor cells to a complete depletion of all stem cells. We showed that committed cells which can divide and are under appropriate crowding feedback control (that is, meeting the sufficient conditions \eqref{eqn:sufficientCond_noOmega} and \eqref{eqn:asymp-stab_2}), will necessarily, without additional mechanisms or assumptions, reacquire stem cell identity, that is, become self-renewing and are at the apex of the lineage hierarchy. Notably, while this process resembles that of dedifferentiation, it does not involve explicit reprogramming, in that the cell state transitions are reversed. Instead, only the cell fate propensities adjust to the changing environment by balancing proliferation and differentiation as is required for self-renewal. While these are purely theoretical considerations, and such a process has not yet been experimentally found, we predict that it must necessarily occur under the appropriate conditions. This can be measured by assessing the gene expression profiles (e.g. via single-cell RNA sequencing) of cells that `dedifferentiate', i.e. reacquire stemness after depletion of stem cells. Moreover, those considerations yield further, more general insights:
\begin{itemize}
    \item Stem cell identity is neither the property of individual cells nor is it strictly associated with particular cell types or states. Any cell that can divide and differentiate, committed or not, may become a stem cell under appropriate environmental control. 
    \item From the latter follows that stemness is a property determined by the environment, not the cell itself.
    \item `Cell plasticity' might need to be seen in a wider context. Usually, cell plasticity is associated with a change of a cell's type when subjected to environmental cues, which involves a substantial remodelling of the cell's morphology and biochemical state. However, we see that a committed cell may turn into a stem cell simply by adjusting the pace of the cell cycle and differentiation processes to the environment. This may not require substantial changes in the cell's state.

\end{itemize}

This exemplifies that homeostatic control through crowding feedback is not only a way to render homeostasis stable and robust, but also to create stem cell identities as a collective property of the tissue cell population.





\backmatter




\bmhead{Acknowledgments}

We thank Ben MacArthur and Ruben Sanchez-Garcia for valuable discussions.

\section*{Declarations}

PG is supported by an MRC New Investigator Award, Grant number MR/R026610/1. The code generated for numerical computations in the current study is available on Github, https://github.com/cp4u17/Feedback. No other data was generated for this work.

Contributions are as follows: C.P. and P.G. conceptualised the paper, C.P. and P.G. did the mathematical analysis, C.P. did the numerical analysis, P.G. supervised the work.

The authors have no competing interests to declare that are relevant to the content of this article.

\begin{appendices}

\section{Asymptotic stability assessment based on Routh-Hurwitz}\label{appendix:RouthHurwitz}

\subsection{Background}

In control system theory, a commonly used method for assessing the stability of a linear system is the Routh-Hurtwiz (RH) criterion \citep{Franklin2014FeedbackSystems}. It is an algebraic criterion providing a necessary and sufficient condition on the parameters of a dynamic system of arbitrary order to ensure the dynamics are asymptotically stable. In particular, the criterion defines a set of conditions on the coefficients, $p_i$, of the characteristic polynomial, $P(s)$, written as
\begin{equation} \label{eqn:characteristicPolynomial}
P(s) = s^n + \sum_{i = 1}^n p_i s^{n-i} \,,
\end{equation}
in which $n$ corresponds to the dimension of the system. Note that the notation used in this section, based on that from \citep{Franklin2014FeedbackSystems}, is different from that of the main text, where $p_i$ is the polynomial coefficient of $i$th order.

A first result of the RH criterion is that a necessary condition for the dynamical system to be asymptotically stable is that all the coefficients must be positive, that is, 
\begin{equation} \label{eqn:RHNecessary}
p_i > 0, \mbox{ for all } i\,.
\end{equation}
Additional conditions on the polynomial coefficients are added for a necessary and sufficient condition. These conditions are based on Routh's array, written as 
\begin{equation} \label{eqn:RHTable}
\begin{bmatrix}
 1 & p_{2} & p_{4} & ... & 0\\
 p_{1} & p_{3} & ... & &\\
 b_1 & b_2 & ... &  &\\
 c_1 & & & &\\
 ...& & & &
\end{bmatrix}, 
\end{equation}
in which the first two rows contain all the coefficients of the characteristic polynomial, and the following ones are recursively computed as 
\begin{equation} 
b_i = -\dfrac{\det\begin{pmatrix} 1 & p_{2i} \\ p_1 & p_{2i+1}\end{pmatrix}}{p_{1}}\,, 
\end{equation}
\begin{equation} 
c_i = -\dfrac{\det\begin{pmatrix} p_1 & p_{2i+1} \\ b_1 & b_{i}\end{pmatrix}}{b_{1}}\,,
\end{equation}
and so on until a zero is encountered. The RH criterion states that the system is asymptotically stable if and only if the elements in the first column of Routh's array are positive. 

Based on that, it can be easily shown that for a second-order polynomial, the necessary condition \eqref{eqn:RHNecessary} is also sufficient for asymptotic stability (\emph{a.s.}) since $b_1 = p_1 p_2$, which means that 
\begin{equation} \label{eqn:RHNecessarySufficient_2D}
\mbox{The system is a. s. } \iff p_i > 0, \mbox{ for } i = 1, 2 \,.
\end{equation}
Instead, the necessary and sufficient condition for a polynomial of order three results in 
\begin{equation} \label{eqn:RHNecessarySufficient_3D}
\mbox{The system is a. s. } \iff p_i > 0, \mbox{ for } i = 1, 2, 3 \, \mbox{ and } p_1 p_2 - p_3 > 0 \,.
\end{equation}
The same reasoning can be applied to higher-order dynamics to derive additional conditions on the coefficients $p_i$.

\subsection{Verification of the necessary condition for asymptotic stability}
\label{app:VerifNecessary}

The Matlab code for verifying \eqref{eqn:c0_mudot1} is provided in https://github.com/ cp4u17/Feedback.git. 

The strategy used is to evaluate each term in Eq. \eqref{eqn:c0_mudot1} and simply compare the left and right-hand sides of the equation. We followed a symbolic approach (based on the Matlab symbolic toolbox) for an arbitrary three-state model. A numerical approach was used instead for higher-order dynamics, specifically 4, 5 and 6 state cell fate models. To do so, we randomly defined the cell dynamical matrix at the steady state, $A(\rho^*)$, and its derivative with respect to $\rho$. Entries were chosen from a uniform distribution and, for assuring a zero dominant eigenvalue for $A(\rho^*)$, a local optimiser (\textit{fmincon} function of Matlab) was used to find appropriate diagonal elements. 
For each dimension of the cell fate model, we tested up to 1000 random cases.

\subsection{Sufficient condition for asymptotic stability}

In this section, we will indicate with the superscripts $A$ and $J$ the coefficients of the characteristic polynomial expressed as Eq. \eqref{eqn:characteristicPolynomial} respectively of the matrix of the dynamical system, Eq. \eqref{eq:AdjacencyMatrixDef}, and those of the Jacobian matrix, Eq. \eqref{eqn:Jacobian_aij}.

For a two and three-state system, the following relations can be algebraically derived
\begin{equation} \label{eqn:coeffPJ2}
p^J_1 = p^A_1 - \sum_i \eta_i\,.
\end{equation}
where $\eta_i$ is according to Eq. \eqref{eq:eta}. Again, considering that $p^A_1 > 0$, if all $\eta_i \leq 0$ then $p^J_1 > 0$. 

Hence, the above relation implies that in a two-state system, the RH criterion given by Eq. \eqref{eqn:RHNecessarySufficient_2D} is fulfilled when $\boldsymbol \eta \leq 0$, with at least one negative component (otherwise $J = A$) and therefore the system is asymptotically stable. We recall that asking $\eta_i \leq 0$ without further constraints is equivalent to the previously derived condition \eqref{eqn:asymp-stab_2} with $\epsilon _i= \infty$.

For applying the RH criterion to a three-state cell dynamic system, given by Eq. \eqref{eqn:RHNecessarySufficient_3D}, we need to evaluate the sign of $p^J_2$ and then that of $p^J_1p^J_2 - p^J_3$. To do so, we first write 
\begin{equation}\label{eqn:PJ3_coeff1}
p^J_2 = p^A_2 - \sum_i f_i \eta_i\,,
\end{equation} 
in which 
$f_i = \sum_j a_{ji} - Tr(A)$ for $i = 1, 2, 3$.
Since the off-diagonal elements are non-negative, and the trace of $A$ is negative, then $f_i > 0$ for $i = 1, 2, 3$. That means that if all $\eta_i \leq 0$ then $p^J_2 > 0$. Concerning the term $p^J_1p^J_2 - p^J_3$, this can be written as a quadratic form in $\boldsymbol \eta = \begin{pmatrix} \eta_1, \eta_2, \eta_3 \end{pmatrix}$ as
\begin{equation} \label{eqn:quadraticEqnEta}
p^J_1p^J_2 - p^J_3 = Q(\boldsymbol \eta) = \boldsymbol \eta^T A_Q \boldsymbol \eta + \bm b_Q^T \boldsymbol \eta+ c_Q\,,
\end{equation}
in which 
\begin{equation}
A_Q = \begin{pmatrix} f_1 & f_1 & f_1 \\ f_2 & f_2 & f_2 \\ f_3 & f_3 & f_3\end{pmatrix} \, , 
\end{equation}
\begin{equation}
\bm b_Q = -p^A_1 \begin{pmatrix} f_1 \\ f_2 \\ f_3 \end{pmatrix} - \dfrac{p_{2}^A}{\boldsymbol v \boldsymbol w}\begin{pmatrix}
v_3 (w_3 - w_1) + v_2 (w_2 - w_1) \\
v_3 (w_3 - w_2) + v_1 (w_1 - w_2) \\ 
v_2 (w_2 - w_3) + v_1 (w_1 - w_3) 
\end{pmatrix} \, ,
\end{equation}
and $c_Q = p_1^Ap_2^A$. Here, $\bm v=(v_1,v_2,v_3)$ is a left dominant eigenvector and $\bm w=(w_1,w_2,w_3)$ a right dominant eigenvector.

We now note that the matrix $A_Q$ is semidefinite positive since two eigenvalues are zero (the rows are two-fold degenerate) and one is positive, equal to $Tr(A_Q) = \sum_i f_i$, and $c_Q>0$. We now distinguish two cases, depending on the sign of $\bm b_Q$ elements. First, if $\bm b_Q \leq 0$, then $Q(\boldsymbol \eta) > 0$ for any $\boldsymbol \eta \leq 0$. Since $f_i,p_1^A,p_2^A, \bm v \bm w > 0$, we get a sufficient condition for $\bm b_Q \leq 0$, namely,
\begin{align}
\label{eq:sufficient_cond_eps_inf}
0 & \leq v_3 (w_3 - w_1) + v_2 (w_2 - w_1) \\ \nonumber
0 &\leq v_3 (w_3 - w_2) + v_1 (w_1 - w_2) \\  \nonumber
0 & \leq v_2 (w_2 - w_3) + v_1 (w_1 - w_3) 
\end{align}
In that case, asymptotic stability and thus crowding feedback control is assured for any $\bm \eta <0$, and thus the bound for feedback strength is $\epsilon_i = \infty$ for $i=1,2,3$.

Otherwise, if there is at least one positive element in $\bm b_Q$, then $Q(\boldsymbol \eta) > 0$ only if $\lvert \eta_i \rvert < \epsilon_i$, where $\bm \epsilon = (\epsilon_1,\epsilon_2,\epsilon_3)$ are the absolute values of the solutions to the equation $Q(\boldsymbol \eta) = 0$, that is -- given that $\eta_i$  are negative -- the solution to,
\begin{equation} \label{eqn:quadraticEqnEpsilon}
0 = \boldsymbol \epsilon^T A_Q \boldsymbol \epsilon - \bm b_Q^T \boldsymbol \epsilon + c_Q\, .
\end{equation}
Importantly, we note that the elements of $\bm b_Q$ depend uniquely on the properties of the dynamical system and therefore, they can be determined without requiring the knowledge of the parameter derivatives, i.e. the specific crowding feedback dependencies.


The Matlab code for verifying \eqref{eqn:coeffPJ2}, \eqref{eqn:PJ3_coeff1} and \eqref{eqn:quadraticEqnEta} is provided in https://github.com/cp4u17/Feedback.git.

\section{Test case} \label{appendix:IllustrativeCase}

\subsection{Asymptotic stability} \label{appendix:asymptoticStability}
\begin{figure}[h]
\begin{center}
\includegraphics[width=0.4\columnwidth]{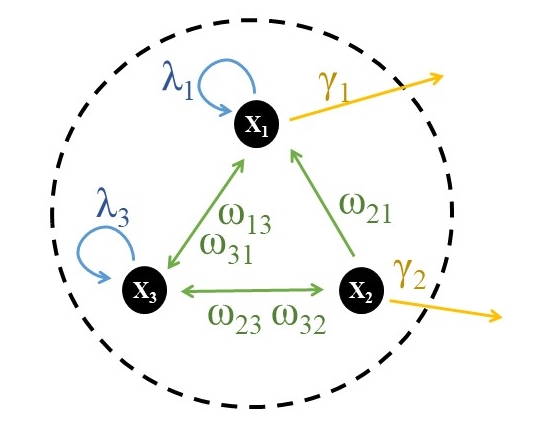} \caption{\label{fig:FeebackTestCases_figNetwork} Cell state network representing a cell type composed of three states. The links represent direct transitions, $\omega_{ij}$; symmetric divisions occur with rates $\lambda_i$ and differentiation with rate $\gamma_i$, where subscripts $i,j=1,2,3$ indicate the corresponding cell state, as per model \eqref{mdl:ExampleModel}.}
\end{center}
\end{figure}

This section reports the details of the model used for numerical examples presented in the main text. The cell dynamics correspond to the following three-state cell fate model
\begin{equation} \label{mdl:ExampleModel}
\begin{aligned}
& X_{1} \xrightarrow{\lambda_{1}} X_{1}+X_{1}, \qquad X_{1} \xrightarrow{\omega_{13}} X_{3}, \qquad X_{1} \xrightarrow{\gamma_{1}} \emptyset \\
& X_{2} \xrightarrow{\omega_{21}} X_{1}, \qquad X_{2} \xrightarrow{\omega_{23}} X_{3}, \qquad X_{2} \xrightarrow{\gamma_{2}} \emptyset \\
& X_{3} \xrightarrow{\lambda_{3}} X_{3}+X_{3}, \qquad X_{3} \xrightarrow{\omega_{31}} X_{1}, \qquad X_{3} \xrightarrow{\omega_{32}} X_{2},
\end{aligned}
\end{equation}
whose network is shown in \fref{fig:FeebackTestCases_figNetwork}. In such a model, for simplicity, we only consider symmetric self-renewing divisions so that $\kappa_{ij} = \omega_{ij}$. Also, we apply the crowding feedback to division rates, $\lambda_i$, and differentiation rates $\gamma_i$. In this way, it is straightforward to apply the sufficient condition \eqref{eqn:asymp-stab_2} for asymptotic stability since $\kappa_{ij}' = 0$ for all $i,j$. 

Hence, each kinetic parameter of the type $\alpha_i \in \{\lambda_j,\gamma_j\}_{j=1,...,3}$ is expressed as a function of $\rho$, whilst those of the type $\alpha_i \in \{\kappa_{jk}\}_{j,k=1,...,3}$ are constant. In particular, we chose a Hill function \citep{Lei2014} where $\alpha_i(\rho) = c_i + k_i \rho^{n_i}/(K_i^{n_i} + \rho^{n_i})$ in case $\alpha_i$ is a differentiation rate, so that $\alpha_i' = \partial \alpha_i / \partial \rho > 0$, and $\alpha_i(\rho) = c_i + k_i/(K_i^{n_i} + \rho/^{n_i})$ in case it is a proliferation rate, so that $\alpha_i' < 0$. According to \eqref{eqn:asymp-stab_2} this choice assures that, if there is a value $\rho = \rho^*$ for which $\mu(\rho^*) = 0$, this corresponds to an asymptotically stable steady state. 

\begin{figure}[h]
\begin{center}
\includegraphics[width=0.45\columnwidth]{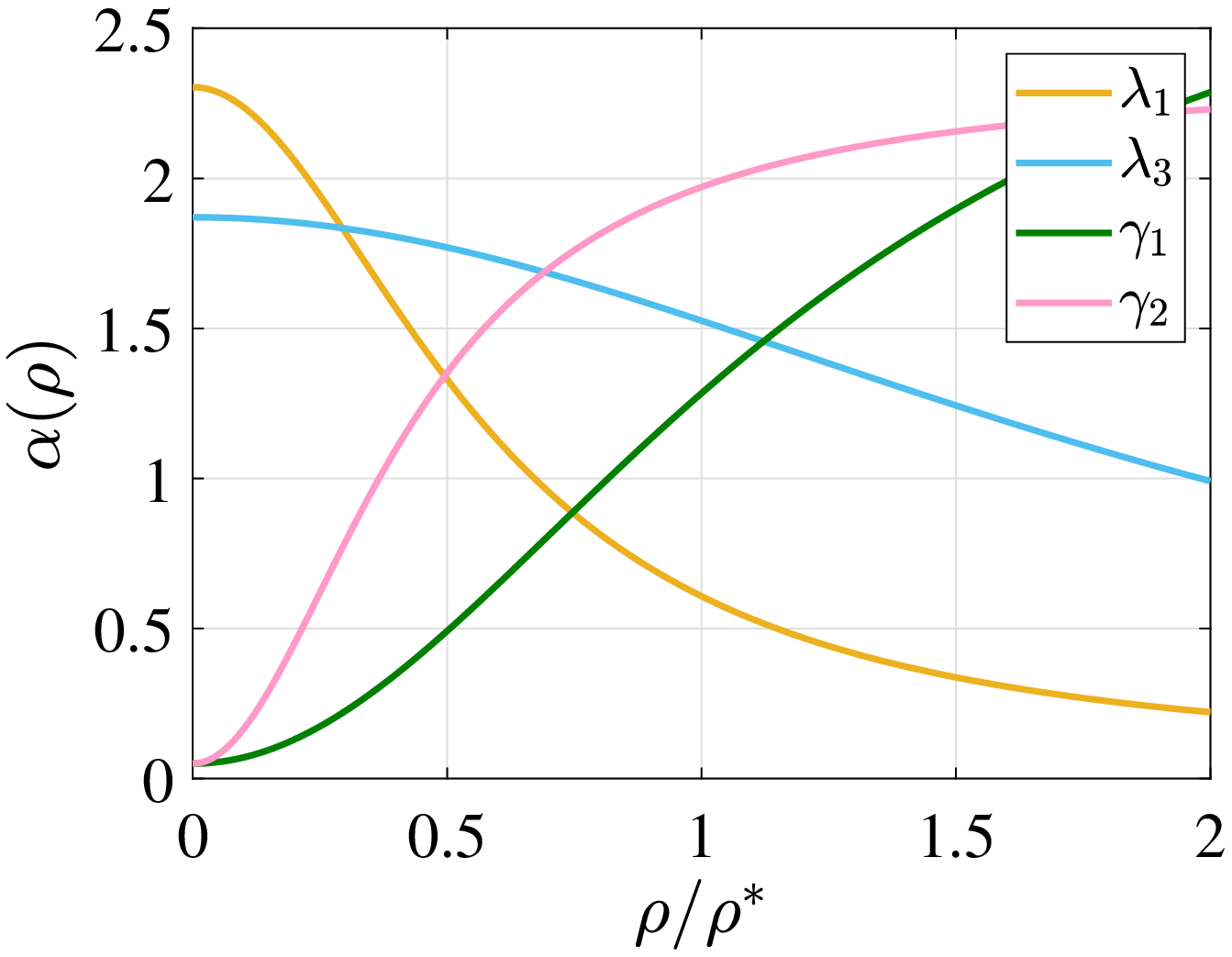} 
\includegraphics[width=0.45\columnwidth]{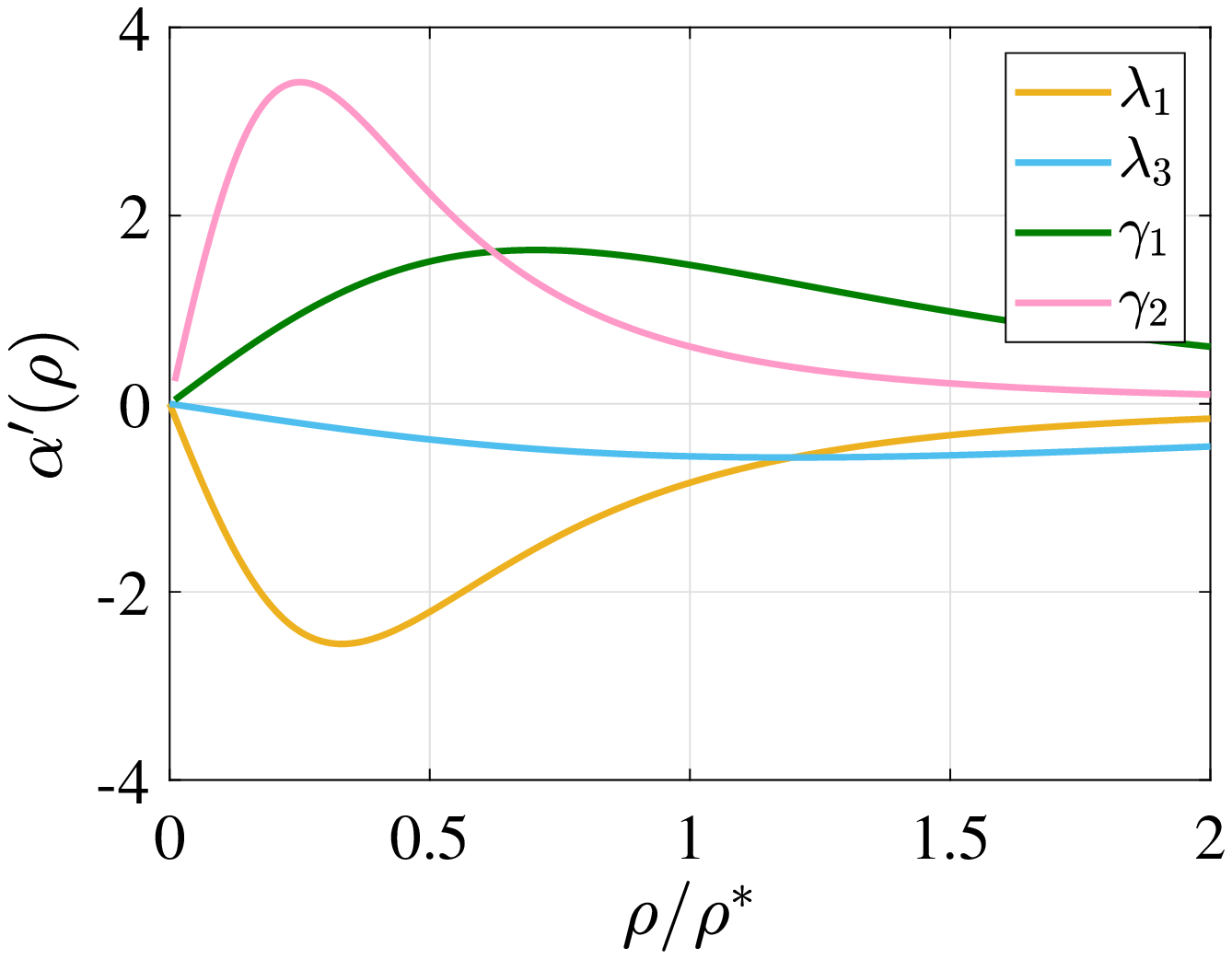} 
\caption{\label{fig:FeebackTestCases_HillFunction}
Proliferation and differentiation rates (left panels, with $\alpha$ as a generic placeholder for parameters), and their derivative with respect to $\rho$ (right panels) as functions of cell density normalised by the steady-state $\rho^*$ for the cell fate model \eqref{mdl:ExampleModel} schematised in \fref{fig:FeebackTestCases_figNetwork}. The profiles in the left panel correspond to Hill functions defined in \tref{tab:FeebackTestCases_table}.
}
\end{center}
\end{figure}

The parameter values used in our example are reported in \tref{tab:FeebackTestCases_table}, and the profiles of the proliferation and differentiation rates as a function of $\rho$ are shown in \fref{fig:FeebackTestCases_HillFunction}. Based on these values, the steady state corresponds to $\rho^* = 1$ (arbitrary unit). As expected, the dominant eigenvalue of the Jacobian at the steady state is negative ($\mu_J = -1.21$).

\begin{figure}[h]
\begin{center}
\includegraphics[width=0.45\columnwidth]{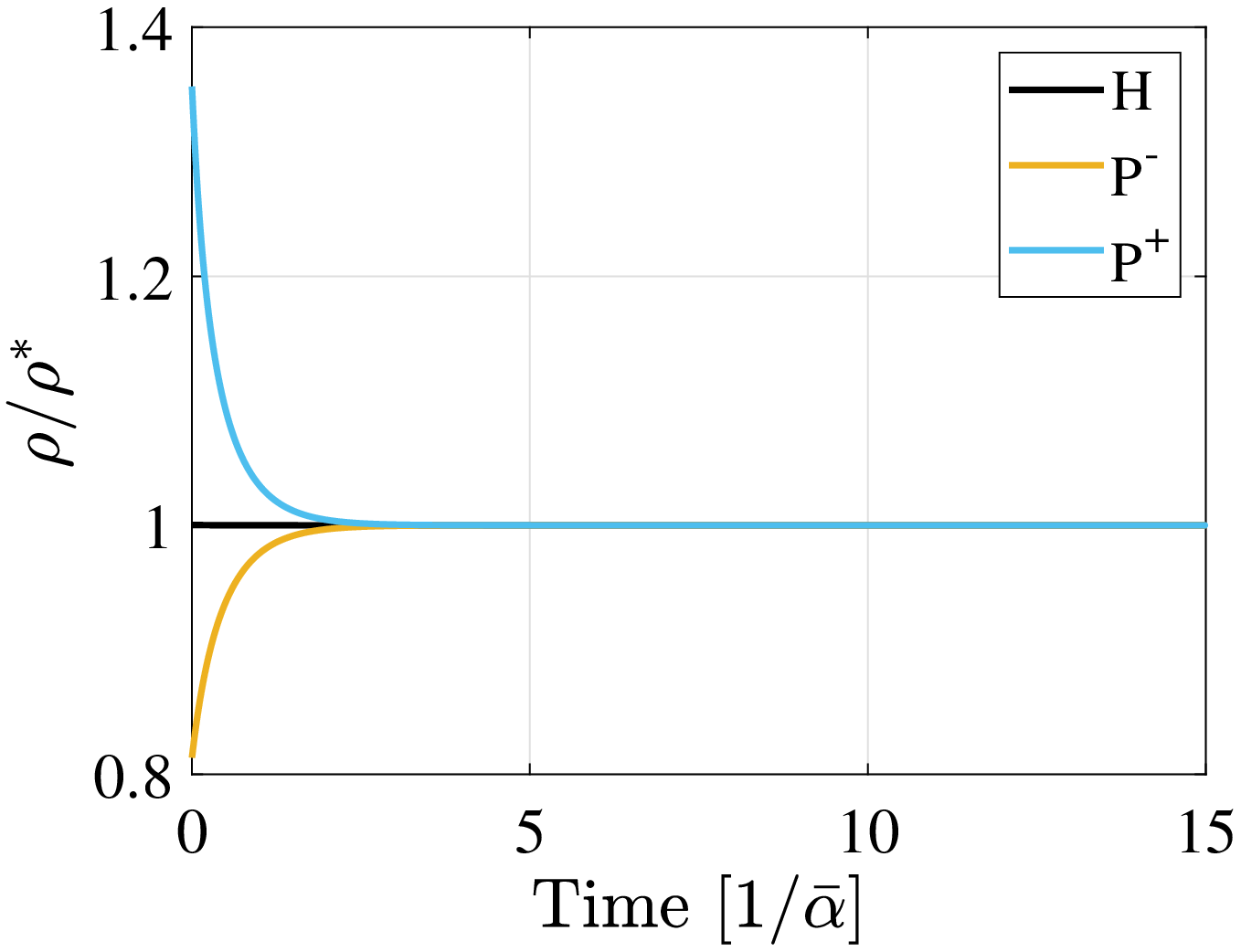} 
\includegraphics[width=0.45\columnwidth]{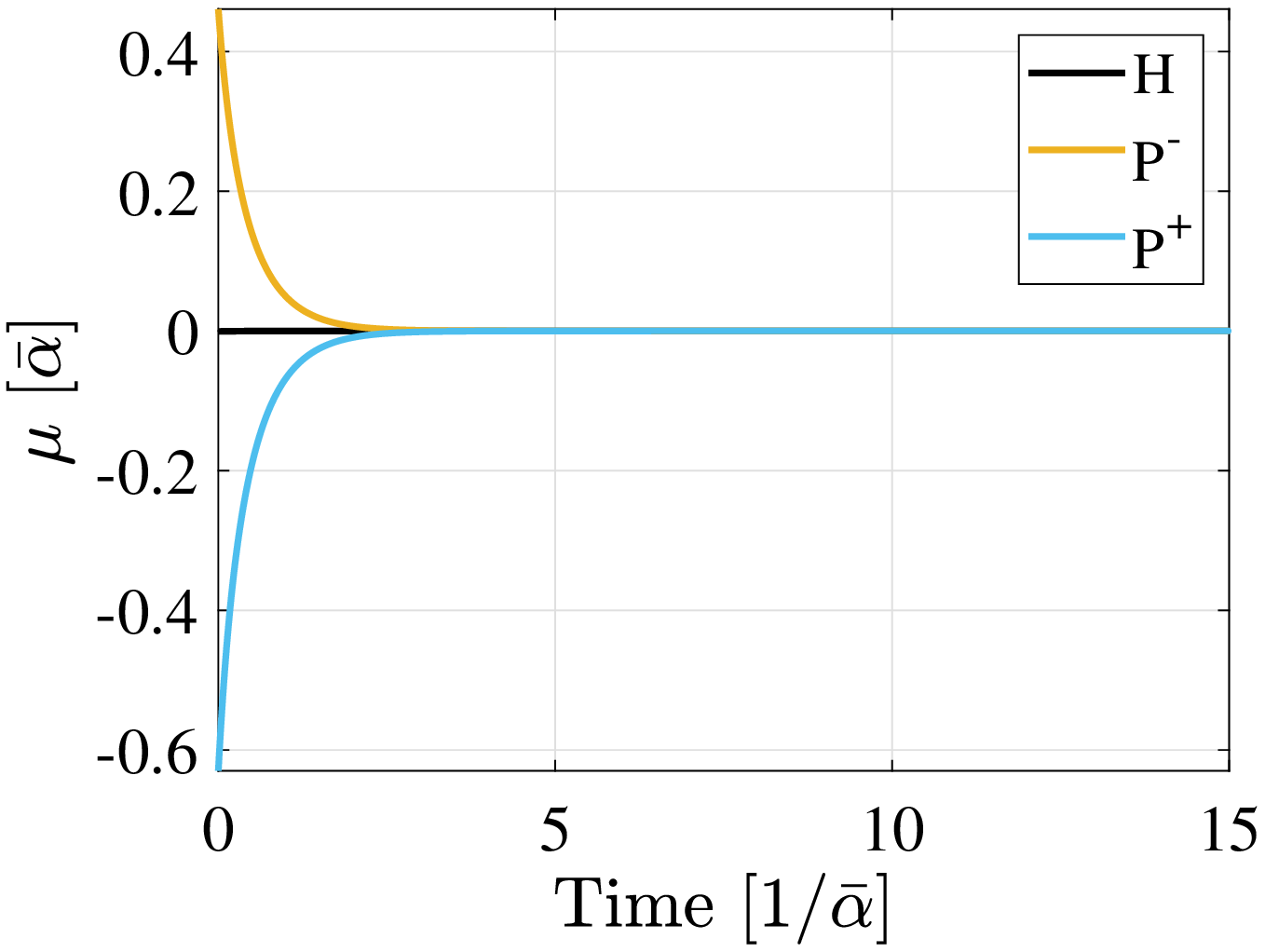}
\caption{\label{fig:FeebackTestCases_Dynamics}
Effect of perturbation of homeostasis under crowding control, when feedback parameters are according to \tref{tab:FeebackTestCases_table}. (Left) Cell density $\rho$, scaled by the steady-state $\rho^*$, as a function of time. (Right) Corresponding variation of the dominant eigenvalue $\mu$. Time is scaled by the inverse of $\bar{\alpha} = \min_i{\alpha_i^*}$. Three different initial condition are tested: \textbf{H}, corresponds to the steady state $\bm \rho^* = (\rho_1^*,\rho_2^*,\rho_3^*)$, \textbf{P}$^-$ to $\begin{pmatrix}0.8 \rho_1^*, 0.75 \rho_2^*, 0.85 \rho_3^* \end{pmatrix}$ and \textbf{P}$^+$ to $\begin{pmatrix}1.5 \rho_1^*, 1.1\rho_1^*, 1.2\rho_1^*\end{pmatrix}$. Since the steady state is asymptotically stable, thanks to crowding control, the cell population remain in, or return to, a homeostatic state characterised by $\mu = 0$.
}
\end{center}
\end{figure}

To test the dynamical behaviour of the tissue cell population, we numerically solved the system of ODEs \eqref{dyn_sys_feedback_eq} for different initial conditions based on the explicit Runge-Kutta Dormand-Prince method (Matlab \textit{ode45} function). The results are shown in \fref{fig:FeebackTestCases_Dynamics} as the time evolution of $\rho$, normalised by the steady-state $\rho^*$, (left panels), and of the dominant eigenvalue, $\mu$ (right panels). The label \textbf{H} indicates an initial condition corresponding to the self-renewing state $\bm \rho^*$, that is, the system is initially in homeostasis. In the simulations labelled as \textbf{P}$^-$ and \textbf{P}$^+$, we applied perturbation in the initial state $\bm \rho^* = (\rho_1^*,\rho_2^*,\rho_3^*)$, which are, respectively, $\begin{pmatrix}0.8 \rho_1^*, 0.75 \rho_2^*,  0.85 \rho_3^* \end{pmatrix}$ and $\begin{pmatrix}1.5 \rho_1^* & 1.1 \rho_2^* & 1.2 \rho_3^* \end{pmatrix}$. As expected, in all these cases, the feedback's effect is stabilising the system so that it returns to the steady state upon perturbation, $\rho \to \rho^*$, (asymptotic stability) and thus regains self-renewal property, $\mu \to 0$, over time. 

\begin{table}[h] \centering
\begin{tabular}{| c| c c c | c c|}
\hline
 & $k$ & $K$ & $n$ & $\alpha^*$ & $\alpha'$\\
\hline
$\lambda_1$ & 0.74 & 0.57 & 2.00 & 0.61 & -0.84\\
$\lambda_3$ & 7.79 & 2.07 & 2.00 & 1.53 & -0.56\\
$\gamma_1$ & 3.07 & 1.22 & 2.00 & 1.28 & 1.48\\
$\gamma_2$ & 2.28 & 0.43 & 2.00 & 1.97 & 0.61\\
\hline
$\kappa_{13}$ & \multicolumn{3}{c|}{--} & 0.95 & 0.00\\
$\kappa_{21}$ & \multicolumn{3}{c|}{--} & 1.44 & 0.00\\
$\kappa_{23}$ & \multicolumn{3}{c|}{--} & 1.71 & 0.00\\
$\kappa_{31}$ & \multicolumn{3}{c|}{--} & 2.03 & 0.00\\
$\kappa_{32}$ & \multicolumn{3}{c|}{--} & 1.35 & 0.00\\
\hline
\end{tabular}
\caption{\label{tab:FeebackTestCases_table}
Values of the Hill function parameters describing the kinetic parameters in case of homeostasis regulation via crowding feedback for the cell fate model \eqref{mdl:ExampleModel}. The generic kinetic parameters (represented as $\alpha_i$ in the right columns of the table) are a function of the total cell density, $\rho$, and are given by $\gamma_i(\rho) = c + k \rho^{n}/(K^{n} + \rho^{n})$ and $\lambda_i(\rho) = c + k/(K^{n} + \rho^{n})$ with $i=1,2,3$. A common value $c = 0.05$ is assumed. State transition rates $\omega_{ij}$, are constant and equal to $\kappa_{ij}$. For such a cell fate dynamics, the steady state is $\rho^* = 1$.
The unit of the kinetic parameter is arbitrary and therefore omitted. Unless specified otherwise, these values apply to all the numerical examples presented in this work.}
\end{table}

\subsection{Failure of feedback function} \label{appendix:FeedbackFailure}

Based on the cell fate model regulated via crowding feedback described in the previous section, we assess the impact of failure in one or more feedback functions. In particular, the failure of the crowding regulation is modelled, assuming one or more kinetic parameters as a constant. Five different failure test cases are assessed. For doing so, we chose $\alpha_i = (1+C) \alpha_i^*$ being constant instead of depending on $\rho$, in which $\alpha^*$ is the value at the steady state when there are no failures (reported in \tref{tab:FeebackTestCases_table}) and $C$ is a constant (reported in \tref{tab:FeebackTestCases_ff_table}). Five test cases, indicated as \textbf{F}$_{1-5}$, are assessed.

\begin{figure}[h]
\begin{center}
\includegraphics[width=0.45\columnwidth]{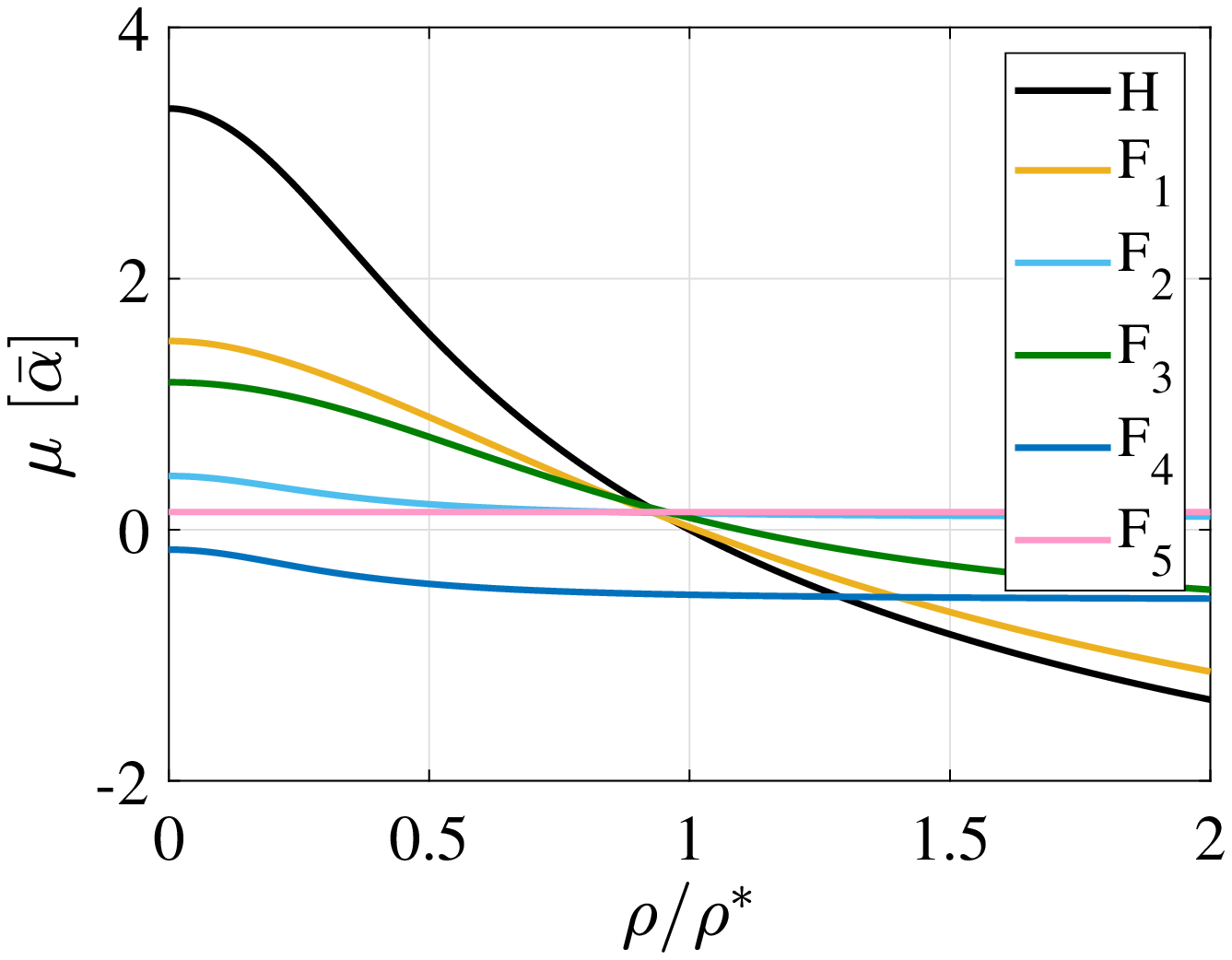}
\caption{\label{fig:FeebackTestCases_FeedbackFailure_eigrho}
Variation of the dominant eigenvalue $\mu$ as a function of the cell density, $\rho$, normalised by the reference homeostatic state value, $\rho^*$. The curve \textbf{H} corresponds to the reference homeostatic model presented in \aref{appendix:asymptoticStability}. The other curves, F$_{1-5}$, represent different sets of feedback failure, as reported in \tref{tab:FeebackTestCases_ff_table}.}
\end{center}
\end{figure}

In test case \textbf{F}$_1$, only one feedback fails. Three of the four kinetic parameters fail in cases \textbf{F}$_{2-4}$. Finally, \textbf{F}$_5$ represents a case where all the feedback functions fail. The corresponding variability of the dominant eigenvalue, $\mu$, as a function of the cell density is shown in \fref{fig:FeebackTestCases_FeedbackFailure_eigrho}. It is clear that whilst \textbf{F}$_{1-4}$ cases satisfy the sufficient condition for strict homeostasis, \eqref{eqn:asymp-stab_2}, in test cases \textbf{F}$_{5}$, the dominant eigenvalue being constant means that there is no homeostatic regulation. Importantly, there is no steady state in test cases \textbf{F}$_{2,4}$ since the dominant eigenvalue is always positive in one case or negative in the other. 

\begin{table}[h] \centering
\begin{tabular}{| c| c c c c c|}
\hline
Parameter & \textbf{F}$_1$ & \textbf{F}$_2$ & \textbf{F}$_3$ & \textbf{F}$_4$ & \textbf{F}$_5$\\
\hline
$\lambda_1$ & +5$\%$ & +5$\%$ & +5$\%$ & -20$\%$ & -5$\%$ \\
$\lambda_3$ & -      & +5$\%$ & +5$\%$ & -20$\%$ & -5$\%$ \\
$\gamma_1$ & -       & -5$\%$ &  -     & +20$\%$ & -5$\%$ \\
$\gamma_2$ & -       & -      & -5$\%$ & -       & -5$\%$ \\
\hline
\end{tabular}
\caption{\label{tab:FeebackTestCases_ff_table} Value of the constant $C$ in the feedback failure test cases. Whenever a failure in the feedback of one kinetic parameter $\alpha$ occurs, that parameter is modelled as a constant, $\alpha = (1+C)\alpha^*$, in which the steady-state value, $\alpha^*$, is reported in \tref{tab:FeebackTestCases_table}. Test cases F$_1$ and F$_2$ correspond to those presented in the main text (\fref{fig:FeedbackFailureTissue}).}
\end{table}

\begin{figure}[h]
\begin{center}
\includegraphics[width=0.45\columnwidth]{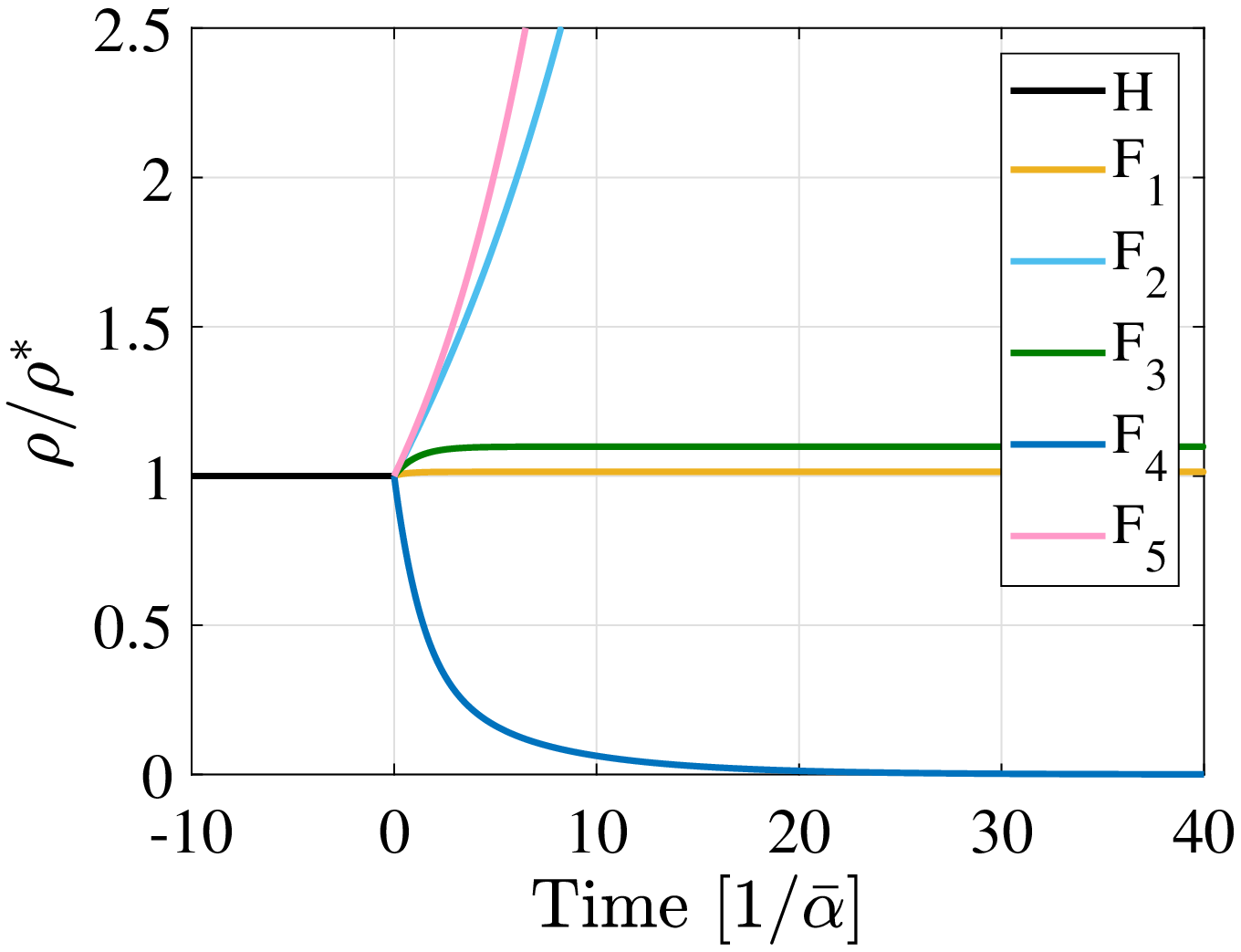} 
\includegraphics[width=0.45\columnwidth]{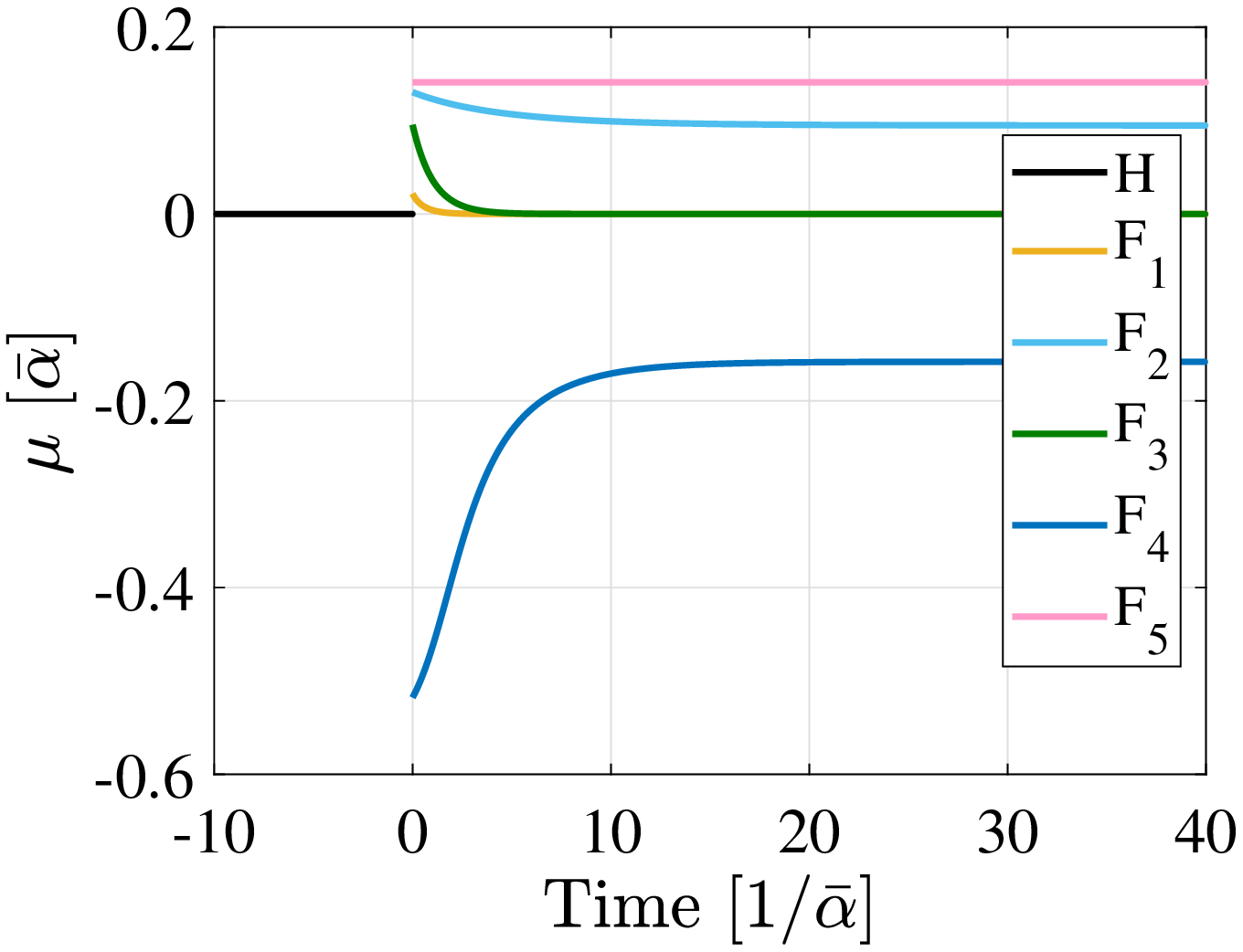}
\caption{\label{fig:FeebackTestCases_FeedbackFailure}
Failure of feedback control. (Left) Cell density, scaled by the steady state in the homeostatic case, as a function of time. (Right) Corresponding variation of the dominant eigenvalue $\mu$. Time is scaled by the inverse of $\bar{\alpha} = \min_i{\alpha_i^*}$. The homeostatic model, \textbf{H}, is perturbed at a time equal to zero to include the feedback failure reported in \tref{tab:FeebackTestCases_ff_table}. Whilst in \textbf{F}$_{1,3}$, the regulation is able to achieve and maintain a new homeostatic state ($\mu = 0$), the remaining case fails to regulate the cell population, leading to an indefinite growth or shrinking of the tissue.}
\end{center}
\end{figure}

Based on these assumptions, we numerically solved the system of ODEs \eqref{dyn_sys_feedback_eq} using the explicit Runge-Kutta Dormand-Prince method (Matlab \textit{ode45} function). The failure test cases start at time 0 from an initially homeostatic condition, \textbf{H}. The results are shown in \fref{fig:FeebackTestCases_FeedbackFailure} as the time evolution of $\rho$, normalised by the homeostatic steady-state, $\rho^*$, (left panels), and of the dominant eigenvalue, $\mu$, (right panels). Note that the cases \textbf{F}$_{1,2}$ correspond respectively to the \textbf{Single failure} and \textbf{Multiple failures} reported in the main text (\fref{fig:FeedbackFailureTissue}).

In two cases, \textbf{F}$_{1,3}$, despite a single or multiple feedback functions failing, a new homeostatic condition is reached after some time, where $\mu = 0$. However, suppose a different set of feedback fails, like in \textbf{F}$_{2,4}$, such that the dominant eigenvalue is respectively positive or negative for any $\rho$. In that case, no steady state can be attained, and the tissue cell population will hyper-proliferate or decline in the long term. Hence, even if the condition for asymptotic stability is met, there is no steady state. Finally, if homeostasis is not regulated at all, as in \textbf{F}$_{5}$, then the population dynamics only depend on the value of the dominant eigenvalue (the cell dynamical model \eqref{dyn_sys_feedback_eq} turns linear). In the case shown, $\mu > 0$ and therefore, the cell population diverges.

\subsection{Single cell mutation scenario} \label{appendix:singleCellMutation}

To assess the tissue dynamics with a single-cell mutation, as presented in the main text, we modelled the clonal dynamics, namely, the dynamics of single cells and their progeny. For doing so, we considered the model \eqref{mdl:ExampleModel} as a Markov process with the same numerical rates as before, but now events are treated as stochastic. Then, we run numerical simulations using the Gillespie algorithm \citep{Gillespie1977} to evaluate this model. In particular, the results presented in this work are based on $100$ independent instances, where each instance is a possible realisation of the stochastic process. We chose a total cell number $N_0 = 5000$ as the initial condition (cell density is based on unitary volume). In real tissues, the number of cells could be a few orders of magnitude larger. However, this number is sufficiently large to avoid the extinction of the process in the time scale analysed, so once rescaled, these dynamics are representative of those in the tissue. All the simulations are stopped when the mutated clone goes extinct or divergence of the dynamics is detected, defined as reaching $N=5 N_0$.

From an implementation point of view, we consider a cell fate model represented by two disconnected cell state networks to model the tissue dynamics, including the mutated cell. One network corresponds to the unperturbed test case \textbf{H}, and the other to the dysregulated one, \textbf{F$_2$} (both described in \aref{appendix:FeedbackFailure}). The simulation starts with $N_0$ cells in the \textbf{H} network, distributed in each state proportionally to the expected steady-state distribution in the tissue, and no cells in the \textbf{F$_2$} network. Thus, since the two networks are disconnected, \textbf{F$_2$} remains empty, and the simulation represents the tissue dynamics before the dysregulation. At a time equal to zero, we moved one cell from a random state in the \textbf{H} network to the corresponding state in the \textbf{F$_2$} one. This simulation represents the tissue dynamics, including the single mutated cell.

\begin{figure}
\begin{center}
\includegraphics[width=0.45\columnwidth]{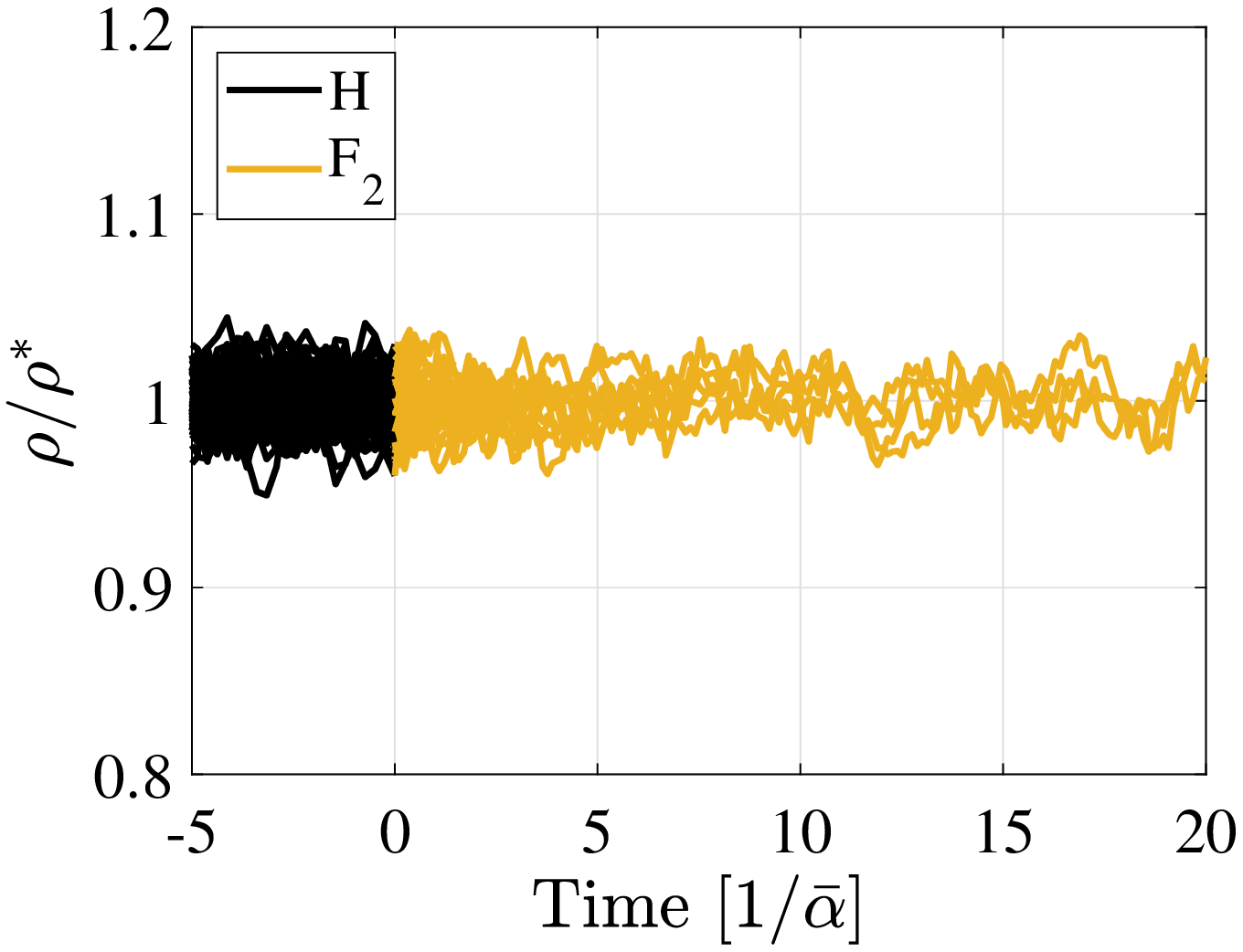}
\includegraphics[width=0.45\columnwidth]{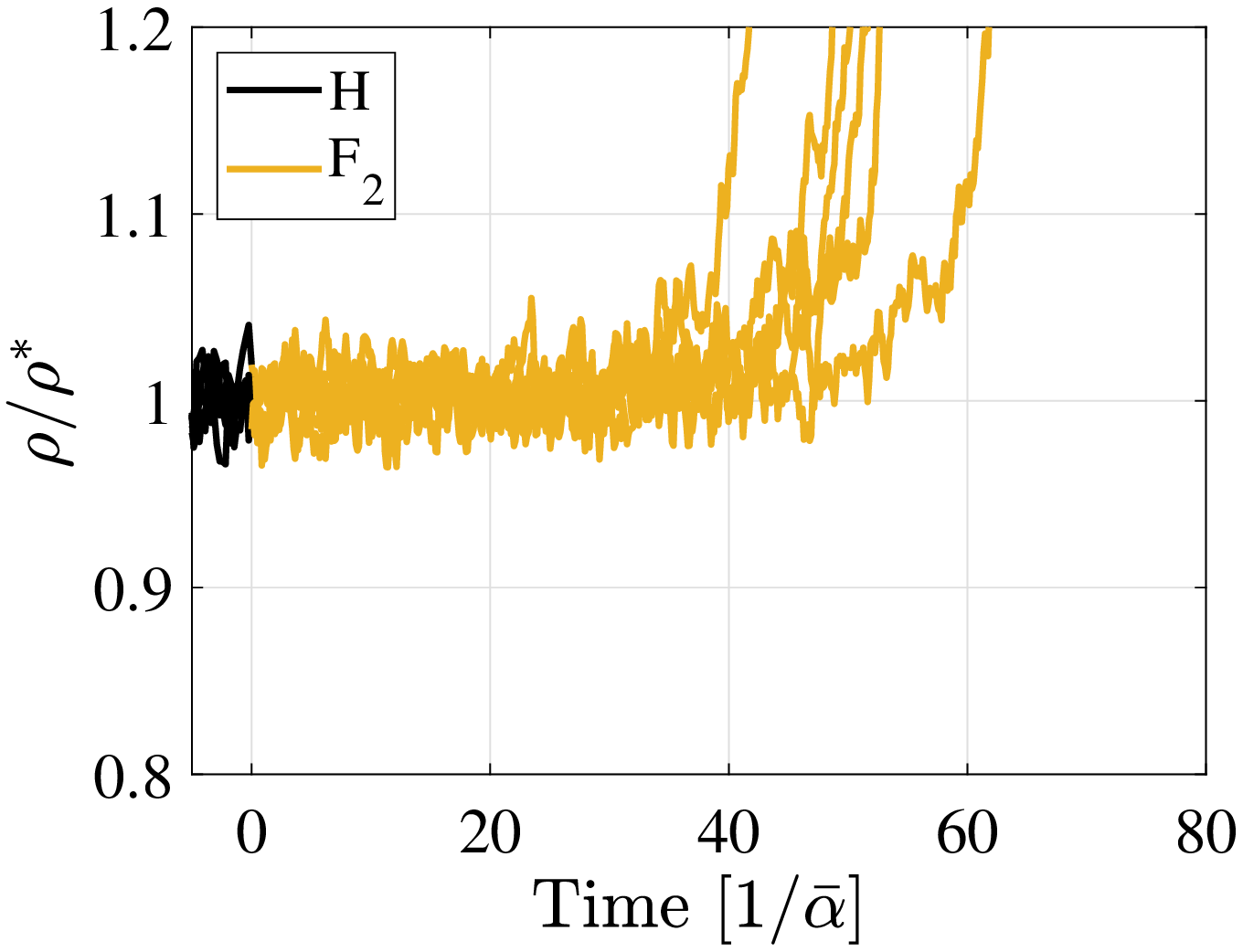}
\caption{\label{fig:TestCaseSingelCellMutation} 
Results of numerical simulations of the stochastic process representing the cell dynamics, according to section \ref{appendix:singleCellMutation}. The cell density, scaled by the steady state in the homeostatic case, as a function of the time is shown for 100 random instances. Each shown trajectory is the result of a different instance of the stochastic process. At a time equal to zero, the cell mutation is modelled as a switch of a single random cell from the homeostatic \textbf{H} cell dynamics to the \textbf{F}$_2$ model assessed in \aref{appendix:FeedbackFailure}. On the left panel, only the trajectories for which the mutated clone goes extinct are shown. The right panel shows the trajectories in which the mutated clone prevails. Dynamics are scaled by $\bar \alpha = \min_{i}\{\alpha_i^*\}$.}
\end{center}
\end{figure}

In \fref{fig:TestCaseSingelCellMutation} (left), all the trajectories where the mutated clones go extinct are shown. In these cases, the tissue dynamics remain globally unaffected by the mutation. Due to the stochastic nature of the process, mutant clones can go extinct even if the growth parameter is positive. That is, even in cases where divergence would be observed for the tissue-wide disruption. However, this does not occur in all the instances. The right panel of the same figure shows those instances where the mutated clone does not go extinct and eventually prevails, resulting in diverging cell population dynamics. For the chosen parameters, this divergence of the mutated clone is detected in 6$\%$ of all cases. Surprisingly, only a few clones survive despite a proliferative advantage, but this is plausible for a small fitness advantage (For example, in the case of a single state with cell division rate $\lambda$ and loss rate $\gamma$ -- a simple
branching process \citep{Haccou2005} -- the probability for the a mutant with $\mu > 0$, that is, $\lambda > \gamma$, to establish is
$1 - \gamma/\lambda$, which can be very low for $\lambda \approx \gamma$).

In the main text (\fref{fig:SingleCellMutation}), only one profile for each scenario is shown, respectively. They correspond to instance $\#24$ for the homeostatic case and instance $\#43$ for the diverging case.

\subsection{Quasi-dedifferentiation} \label{appendix:qDedifferentiation}

The numerical example presented in the main text is based on the same cell fate model described in \aref{appendix:asymptoticStability}. To model the dynamics of a committed cell type, we choose a constant non-negative $\bm u = \begin{pmatrix}0.02 & 0.07 & 0.06\end{pmatrix}^T$ to model for the cell influx. For such a model, the steady state, $\rho_c^*$, is asymptotically stable. 

The figures presented in the main text are based on the numerical integration of the system of ordinary differential equation \eqref{eqn:dyn_feedback_cellTypeU}. In particular, we used the explicit Runge-Kutta Dormand-Prince method (Matlab \textit{ode45} function).

\end{appendices}

\bibliography{references}

\end{document}